\documentclass[twocolumn]{aastex61}
\pdfoutput=1
\usepackage{graphicx}
\usepackage{amsmath}
\usepackage{epsfig}
\usepackage{bm}

\newcommand \hii {H{\small II}\ }

\newcommand \kms {km s$^{-1} \ $}
\newcommand \as  {$^{\prime\prime}~$}
\newcommand \am  {$^{\prime}~$}

\shorttitle{W51C}
\shortauthors{M.F.Zhang et al.}
\begin{document}
\title{Disentangling the radio emission of the supernova remnant W51C}

\author{M.F.Zhang}
\affiliation{Key Laboratory of Optical Astronomy, National Astronomical Observatories, Chinese Academy of Sciences,
Beijing 100012, China}
\affiliation{University of Chinese Academy of Sciences, 19A Yuquan Road, Shijingshan District, Beijing 100049, China}

\author{W.W.Tian}
\affiliation{Key Laboratory of Optical Astronomy, National Astronomical Observatories, Chinese Academy of Sciences,
Beijing 100012, China}
\affiliation{University of Chinese Academy of Sciences, 19A Yuquan Road, Shijingshan District, Beijing 100049, China}
\affiliation{Department of Physics $\&$ Astronomy, University of Calgary, Calgary, Alberta T2N 1N4, Canada}

\author{D.A.Leahy}
\affiliation{Department of Physics $\&$ Astronomy, University of Calgary, Calgary, Alberta T2N 1N4, Canada}

\author{H.Zhu}
\affiliation{Key Laboratory of Optical Astronomy, National Astronomical Observatories, Chinese Academy of Sciences,
Beijing 100012, China}
\affiliation{Harvard-Smithsonian Center for Astrophysics, 60 Garden Street, Cambridge, MA 02138, USA}

\author{X.H.Cui}
\affiliation{Key Laboratory of Optical Astronomy, National Astronomical Observatories, Chinese Academy of Sciences,
Beijing 100012, China}

\author{S.S.Shan}
\affiliation{Key Laboratory of Optical Astronomy, National Astronomical Observatories, Chinese Academy of Sciences,
Beijing 100012, China}
\affiliation{University of Chinese Academy of Sciences, 19A Yuquan Road, Shijingshan District, Beijing 100049, China}

\correspondingauthor{W.W.Tian}
\email{tww@bao.ac.cn}

% \author[0000-0001-8261-3254]{
% M.F.Zhang\altaffilmark{1,2},
% W.W.Tian\altaffilmark{1},
% D.A.Leahy\altaffilmark{3},
% H.Zhu\altaffilmark{1,4},
% X.H.Cui\altaffilmark{1}
% }
% \affil{
% \sp{1} Key Laboratory of Optical Astronomy, National Astronomical Observatories, Chinese Academy of Sciences,
% Beijing 100012, China \\
% \sp{2} University of Chinese Academy of Sciences, 19A Yuquan Road, Shijingshan District, Beijing 100049, China\\
% \sp{3} Department of Physics $\&$ Astronomy, University of Calgary, Calgary, Alberta T2N 1N4, Canada\\
% \sp{4} Harvard-Smithsonian Center for Astrophysics, 60 Garden Street, Cambridge, MA 02138, USA\\
% }

% \email{tww@bao.ac.cn}

% \maketitle
\label{firstpage}
\begin{abstract}
We simulate the evolution of supernova remnant (SNR) W51C.
The simulation shows the existence of a new northeast edge.
We present magnetic field structure of the W51 complex (SNR W51C and two \hii regions W51A/B) by employing the 11 cm
survey data of Effelsberg.
This new edge is identified and overlaps with W51A along the line of sight, which gives a new angular diameter of about
37\am for the quasi-circular remnant.
In addition, we assemble the OH spectral lines (1612/1665/1720 MHz) towards the complex by employing the newly
released \textit{THOR} (The HI OH Recombination line survey of Milky Way) data.
We find that the known 1720 MHz OH maser in the W51B/C overlap area is located away from the detected 1612/1665MHz
absorption region.
The latter is sitting at the peak of the \hii region G49.2-0.35 within W51B.
\end{abstract}


\section{Introduction}
The interaction between supernova remnants (SNRs) and molecular clouds (MCs) has impact on star formation,
the origin of cosmic rays, and the Galactic interstellar medium (ISM).
The shocks of supernovae can compress MCs, and could be a good catalyst for star formation.
The shock can also dissociate molecules and generate high-energy particles, which are an important
source of Galactic cosmic rays.
Most of heavy elements are produced through supernova explosions, while the interaction between SNR shocks and MCs
plays a key role on synthesis of various ISM molecules.
A detailed study on SNR-MCs interaction can help understand these physical processes better.

The W51 complex is composed of two \hii regions (W51A/B) and one SNR (W51C).
Both \hii regions are extended and consist of many small \hii structures, such as G49.2-0.35.
W51C has a thick semicircular shell of 14\am $\times$ 20\am \citep{Copetti1991,Subrahmanyan1995}
in the radio continuum image, one side of which adjoins the \hii region W51B.
Recent observations show high-energy features in the boundary \citep{Abdo2009,Aleksic2012},
which are regarded as a strong evidence of interaction between the SNR and MCs.
In the east of W51B, i.e., next to the SNR explosion center, 1720 MHz OH masers were discovered \citep{Hewitt2008,Brogan2013},
which is a further evidence supporting an interaction.
The W51A is an active star formation region, but does not show any sign of connection with W51C.
W51A and W51B are connected along the line of sight (LoS) in CO and infrared images \citep{Kang2010,Parsons2012,Ginsburg2015}.
They have distances of 5 kpc $\sim$ 8 kpc \citep{Genzel1981,Schneps1981,Xu2009,Sato2010,Tian2013}.
For comparison, the distance of SNR W51C is 4.3 kpc \citep{Tian2013} to 6 kpc \citep{Koo1995}.
The spatial location of the three sources seems hint a physical connection among them.

Many observations of the complex have been done so far, but the physical association among the major structures of the complex
is still not clear.
Thus, we create a MHD simulation of the SNR in order to study the possible physical connection between the remnant and
the \hii regions.
In addition, we analyze the polarization data and the OH spectra data to understand the simulation results.

In Sect.2, we describe the data and processing method.
In Sect.3, we give the simulation model and list the parameters we use.
In Sect.4, we present the results.
In Sect.5, we discuss the results.
Sect.6 is a summary.

\section{Data Reduction}
The data used for generating the magnetic field structure is from the Effelsberg 11cm (2.695GHz) survey
\footnote{http://www3.mpifr-bonn.mpg.de/survey.html} \citep{1999A&A...350..447D}.
\citet{1999A&A...350..447D} presented a polarization map covering the W51 region.
They smoothed the map to cover a larger region, which obscured some small structures.
The resolution of the original 11 cm data is 5\am and the resolution of the smoothed images is 12\am.
In Fig.~\ref{fig:mag}, we show the map of W51 at 5\am resolution and rotate the polarization
direction $90^{\circ}$ to get an estimate of the magnetic direction, i.e., the arrows indicate magnetic
direction, while the arrow length indicates polarization intensity.
% Usually, polarization intensity is positively correlated with magnetic intensity, assuming the radio emission
% origins from synchrotron mechanism.
Because the 11cm data have an instrumental polarization of $0.7\% \pm 0.25\%$ \citep{1987A&AS...69..451J}, we have to
subtract the instrumental polarization intensity.
At first, we delete the part of the data with values below one standard deviation in the total intensity image.
Then we obtain the polarization intensity by $I^{final}_{pol}=I^{primary}_{pol}-I_{total}*0.007$ and set negative values to
zero. Here $I^{final}_{pol}$ is the obtained polarization intensity, $I^{primary}_{pol}$ is the original polarization
intensity, and $I_{total}$ is the total continuum intensity.
Finally, we obtain the degree of polarization by $p=I^{final}_{pol}/I_{total}$.
In fact, the instrumental polarization intensity in part area may be less than $0.7\%$, so our subtracted map shows many
negative values.
In our processing, all negative values are set to zero.
The continuum and polarization sensitivity of the survey are 20 mK and 11 mK respectively.

We use the newly-released data from \textit{THOR} DR1 \citep{Beuther2016}.
DR1 consists of the 1.4 GHz continuum maps, HI, OH spectral line maps (1612/1665/1720 MHz) and radio recombination
line maps, covering a region with $15^{\circ}<l<67^{\circ}$,$-1^{\circ}<b<1^{\circ}$.
The spatial resolution of the OH spectral lines is about 20\as and the velocity resolution is about 1.5 \kms.
Because the OH maser lines are so strong that we cannot plot them with other spectral lines at the same scale, we
modify the spectral scale in maser regions, where the spectral intensities are divided by 20.

The 1.4 GHz continuum image is from the \textit{VGPS} (VLA Galactic Plane Survey) \citep{Stil2006}.
The spatial resolution of the \textit{VGPS} is 1\am and the sensitivity is 2 K.

\section{Simulation model}

\begin{table}
  \caption{Summary of Simulation Parameters on W51C}
  \label{table:parameters}
  \centering
  \begin{tabular}{l l l}
      \hline\hline
      Parameters                      & Value           & References               \\
      \hline
      Ejecta Mass                     & 11 M$_{\odot}$ & 1, 2\\
      Initial Explosion Energy        & 1.3$\times$ 10$^{51}$ ergs & 3, 4\\
      Initial Radius                  & 4 pc            &\\
      Initial Time                    & 850 years       & 5\\
      \hline
      Mean Density                    & 0.21\ cm$^{-3}$ & 4, 5\\
      Density Index ($\alpha$)          & 2.4             & 6\\
      Magnetic Field Gradient         & 0.1 $\mu$G pc$^{-1}$    &\\
      Central Magnetic Intensity      & 9 $\mu$G        &\\
      Mean Atomic Weight              & 1.3             &\\
      Adiabatic Coefficient           & 5/3             &\\
      Temperature                     & 100 K           &\\
      \hline
      Synchrotron Index ($\beta$)       & 0.25            & 7\\
      Distance                        & 4.3 kpc         & 8\\
      \hline
  \end{tabular}\\
  \tablerefs{(1)\citealt{Sasaki2014}; (2)\citealt{Sukhbold2016}; (3)\citealt{Poznanski2013}; (4)\citealt{Koo1995};
  (5)\citealt{Leahy2017a}; (6)\citealt{Parsons2012}; (7)\citealt{1970AuJPA..14..133S}; (8)\citealt{Tian2013}}
\end{table}

The simulation is based on a 3D (three-dimensional) MHD (magnetohydrodynamic) model, in which we ignore all dissipative
effects, such as viscosity, resistivity, and thermal conduction.
Gravitation and radiation cooling are also removed from the model, because they have little influence on the
simulation.

In terms of MHD theory, the ideal conservation equation is:
\begin{equation}
    \begin{cases}
      \dfrac{\partial \rho}{\partial t} + \nabla \cdot (\rho \bm{v}) = 0 , \\
      \dfrac{\partial \rho \bm{v}}{\partial t} + \nabla \cdot (\rho \bm{vv} - \bm{BB}) + \nabla P^* = 0 , \\
      \dfrac{\partial E}{\partial t} + \nabla \cdot [(E+P^*)\bm{v} - \bm{B}(\bm{v} \cdot \bm{B})] = 0 , \\
      \dfrac{\partial \bm{B}}{\partial t} + \nabla \times (\bm{v} \times \bm{B}) = 0,
    \end{cases}
\end{equation}
in which, $\rho$ is mass density, $\bm{v}$ is velocity, $\bm{B}$ is magnetic field intensity, $P^*$ is total
pressure, and $E$ is total energy density.

\citet{Sasaki2014} estimated W51C's progenitor mass to be more than 20 M$_\odot$ using a distance of 6 kpc.
This distance is an upper limit, so the real mass is less.
% If we take 4.3 kpc as the distance, the progenitor mass is more than 7.4 $M_\odot$.
In the simulation, we use 20 M$_\odot$ as the progenitor mass, then the ejecta mass should be 11 M$_\odot$
\citep{Sukhbold2016}.
\citet{Koo1995} estimated an explosion energy of 3.6 $\times$ 10$^{51}$ ergs s$^{-1}$, which is larger than the canonical value,
1 $\sim$ 3 $\times$ 10$^{51}$ ergs s$^{-1}$ \citep{Poznanski2013}.
For W51C, the explosion energy should be about 1.0 $\times$ 10$^{51}$ ergs s$^{-1}$ based on the work of \citet{Poznanski2013}.
Although \citet{Poznanski2013} only estimated the kinetic energy of SNRs, those SNRs are so young that we think the
kinetic energy is nearly equal to the total explosion energy.
In contrast, W51C is a middle-aged SNR, thus its initial explosion energy is the sum of the present kinetic energy and thermal
energy.
Using a distance of 6 kpc, \citet{Koo1995} derived the thermal energy and estimated the explosion energy according to Sedov
model \citep{1959sdmm.book.....S}.
Taking the recent measured distance of 4.3 kpc \citep{Tian2013}, we employ same method as \citet{Koo1995} and get an
explosion energy of 1.3 $\times$ 10$^{51}$ ergs s$^{-1}$, more reasonable for such a progenitor and similar to the estimation
based on the work of \citet{Poznanski2013}.
% Thus, in the study, we use 4.3 kpc as the distance of W51C.
Of course, the explosion energies of SNRs vary in a large range \citep{Leahy2017}, so it can not be used to proving which distance
is right.
At least, 4.3 kpc is more reliable in terms of present evidences, so we use it as the distance of W51C.
We take the region of $48.89^{\circ}<l<49.51^{\circ}$,$-0.81^{\circ}<b<-0.19^{\circ}$ as the new size of W51C.
For such a quasi-circular SNR with angular diameter of about 37\am, its shock diameter is 46 pc.
The age is 18000 years, if we use the shock velocity 490 \kms derived by \citet{Koo1995}.

We use a python calculator for simulating SNR evolution \citep{Leahy2017a} to estimate the ambient density of 0.21 cm$^{-3}$.
To simulate a random distribution of density, we adopt a power-law density distribution $N(\rho) = N_0 \rho^{-\alpha}$,
in which $N(\rho)$ is the pixel numbers at different densities, $N_0$ is a constant,
$\alpha$ is the power-law index, which is a positive number, i.e., low-density regions are larger than high-density regions.
For example, there are 256 $\times$ 256 $\times$ 256 pixels in our simulation, then there are possibly 10,000,000 pixels with densities lower
than 0.21 cm$^{-3}$ and only 100 pixels with density higher than 21 cm$^{-3}$.
In the initial conditions, we only warrant the mean density of 0.21 cm$^{-3}$.
\citet{Parsons2012} studied the molecular clumps (including dense clumps) and found their mass distribution follows a
power-law with an index of 2.4, which may imply the distribution of ambient density.
For this region, no atomic hydrogen density distribution is given at present, so we take $\alpha$ = 2.4 in the initial density distribution.
% \textbf{Moreover, we have tested the simulation when $\alpha$ = 1.0, then the radio image becomes more irregular and has many random components.}
In addition, to simulate the interaction, we add a MC with a density of 2 cm$^{-3}$ and a diameter of 11 pc (see two middle panels of
Fig.~\ref{fig:simulation}).

For the magnetic field, we apply a direction of $45^\circ$ rotating anti-clockwise from Galactic plane to reconstruct
the morphology of W51C, because the direction of W51C edge is like this.
We assume that there is same magnetic direction in this region.
We want to reproduce a much lower radio surface brightness for the upper rim of W51C compared to the lower rim, because we want to
get the largest difference between the new edge and the old edge of W51C, so that we can get the smallest radio flux density of the new edge.
Meanwhile, the magnetic intensity should be reasonable.
The reasonable magnetic intensity in Milky Way is from 4 $\mu$G to 14 $\mu$G \citep{Haverkorn2015}. The length range in the simulation is
$75 \times \sqrt{2}$ pc. Then the magnetic gradient is 0.1 $\mu$G pc$^{-1}$ and the central magnetic field is 9 $\mu$G.
% In a l-b (longitude-latitude) coordinate, the intensity of magnetic field decreases from southwest to northeast with
% a gradient of 0.1 $\mu$G pc$^{-1}$.
% To get a large radio flux density difference between two edges, we have to set as large gradient as possible.
% The magnetic intensity of the explosion site is set to 9 $\mu$G.
% Then the largest and the smallest initial magnetic intensities are about 14 $\mu$G and 4 $\mu$G, both reasonable in
% Milky way \citep{Haverkorn2015}.
% Larger gradient will lead to unreliable magnetic intensities.

In the simulation, a region of $1^\circ \times 1^\circ$ is 75 pc $\times$ 75 pc in real space.
For such a region, we generate a grid of 256 $\times$ 256 $\times$ 256, i.e. its resolution is 0.3 pc pixel$^{-1}$
(0.24\am pixel$^{-1}$).
To get an approximate circular explosion and ensure the shock is still in an ejecta-dominated phase at initial time, we
use 4 pc as the initial radius.
For SNR W51C, it would take about 850 years to expand to 4 pc in size.
Meanwhile, the Sedov time of SNR W51C is 3200 years \citep{Leahy2017a}, so it is before the Sedov phase and the setting is appropriate.
In Fig.~\ref{fig:simulation}, the top panels show the initial conditions in the simulation.

To compare with real observation, we use $i(\nu)=C\rho B_{\perp}^{\beta + 1}\nu^{-\beta}$ \citep{Orlando2007}
to get the relative radio flux volume density, in which $\nu$ is the radiation frequency, C is a constant, $\rho$ is the density,
$B_{\perp}$ is the magnetic field perpendicular to the LoS and $\beta$ is the synchrotron spectral index.
By employing $\int i(\nu) \mathrm{d}l$, we integrate $i(\nu)$ along LoS to obtain relative radio flux density.
In this simulation, the x-axis is defined as the direction of LoS.
We assume all radio radiation originates from synchrotron mechanism.
For the interaction region, we think about 10$\%$ of the material contribute the radio flux density, i.e.
$i(\nu)=0.1C\rho B_{\perp}^{\beta + 1}\nu^{-\beta}$.
The $\rho$ and $B_{\perp}$ are derived by simulation, while the $\beta$ is 0.25 for W51C \citep{1970AuJPA..14..133S}.
Then we get a radio image.
The resolution of \textit{VGPS} continuum image is 1\am and the full width at half maximum (FWHM) for a Gaussian function
is $2\sqrt{2ln2} \sigma$, in which the Gaussian function is:
\begin{equation}
  \begin{aligned}
    G(x)=\dfrac{1}{\sigma \sqrt{2\pi}}e^{-\dfrac{1}{2}(\dfrac{x-\mu}{\sigma})^2},
  \end{aligned}
\end{equation}
where $\sigma$ is the standard deviation, $\mu$ is the expectation value.
The resolution is equal to the FWHM, so the $\sigma = 0.42$\am for a beam of \textit{VGPS}.
In our simulation, the resolution is 0.12\am/pixel, so the $\sigma = 3.5$ referring to the number of pixels.
For simplicity, we set $\sigma = 4$.
Thus, we smooth the image using a 2D Gaussian function with $\sigma = 4$.
% A small $\sigma$ will lead to the absence of large-scale features, which is more important for SNR W51C.
% We indeed do not find the thick shell in primary \textit{THOR} continuum data that does not combine the single dish
% data.
% In other words, the shell can only be found in data containing information of large-scale structures.
% To compare with the observations of single-dish telescopes, we also assume $\sigma = 14$ (about 4\am for real observation)
% to simulate the influence.
% Our simulation is in 2D frame, the real radio flux density is $I=\int{i(\nu)}}dl$, so the real shell would be thicker.

The evolutions with different parameters are included in the work.
We take 2.0 and 3.0 $\times$ 10$^{51}$ ergs as possible explosion kinetic energy for W51C,
while keep other parameters same as the previous settings.
We also respectively set 0.13 cm$^{-3}$ and 0.3 cm$^{-3}$ as the mean ISM density of the whole region.

We use PLUTO \footnote{http://plutocode.ph.unito.it/} \citep{Mignone2007} to perform the simulation.
The initial conditions are summarized in Table.~\ref{table:parameters}, in which the parameters without references
are from our estimations.
According to our model, we simulate the evolution of SNR W51C till 18000 years.

\section{Results}
\subsection{Simulation}

\begin{figure*}
    \centering
    \includegraphics[width=0.323\textwidth]{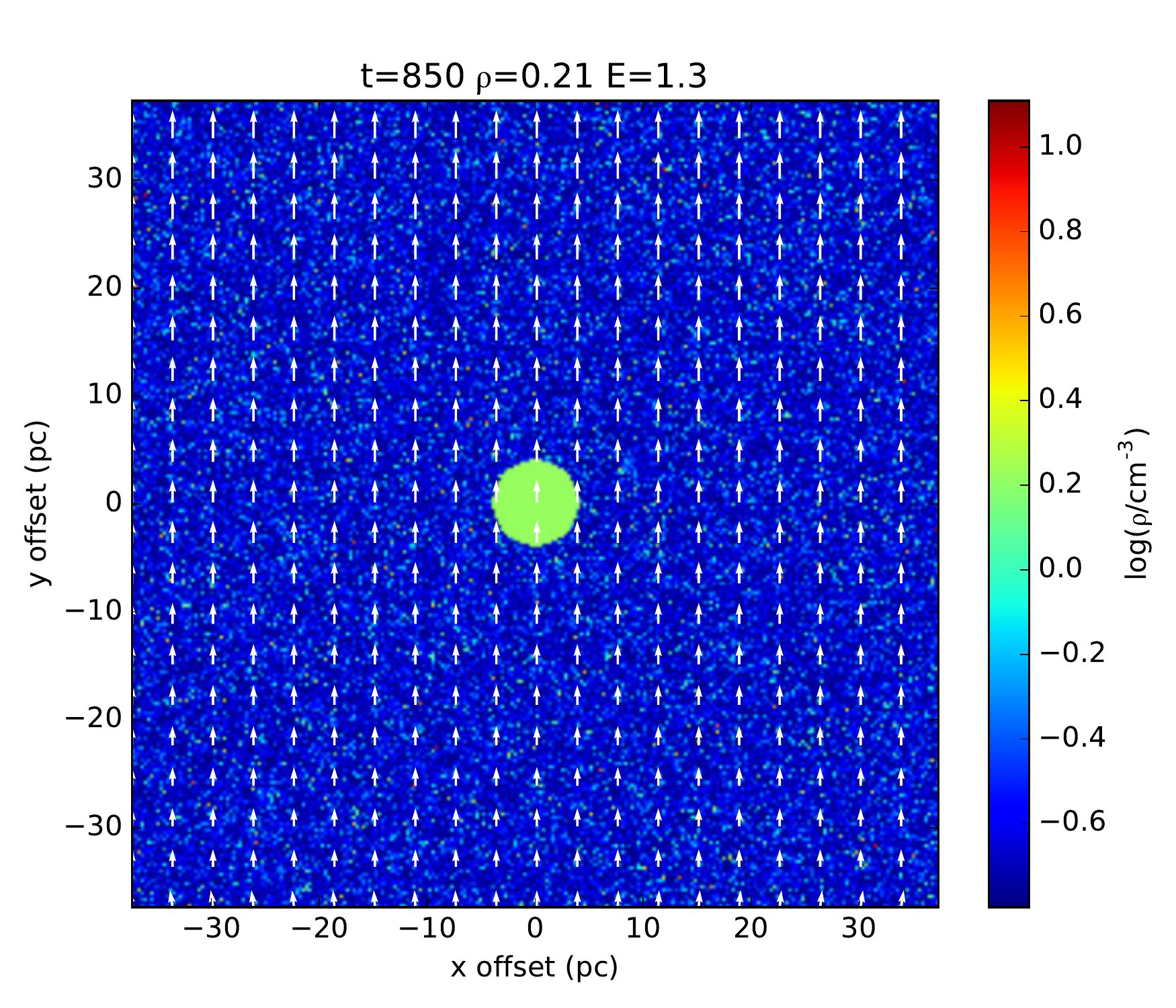}
    \includegraphics[width=0.323\textwidth]{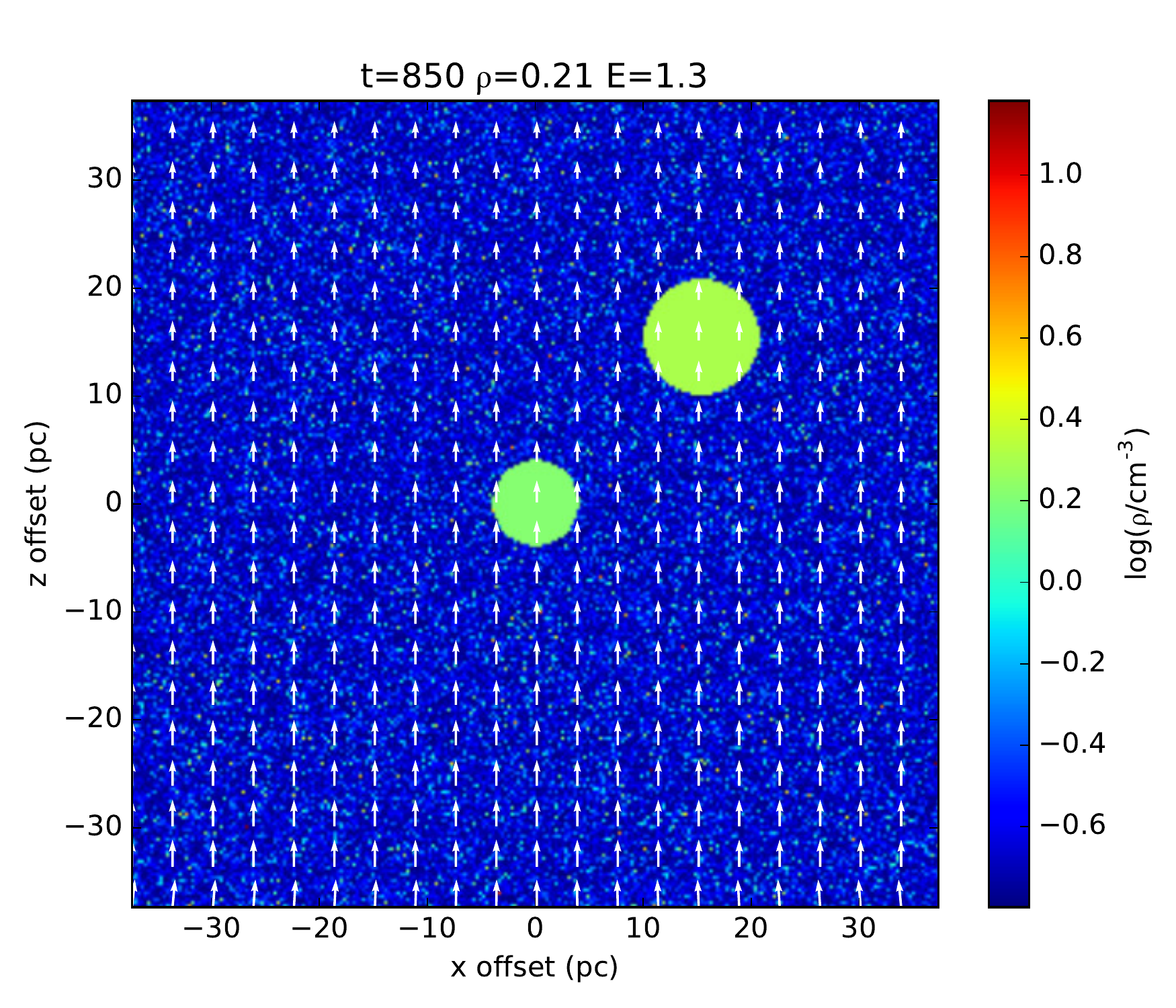}
    \includegraphics[width=0.323\textwidth]{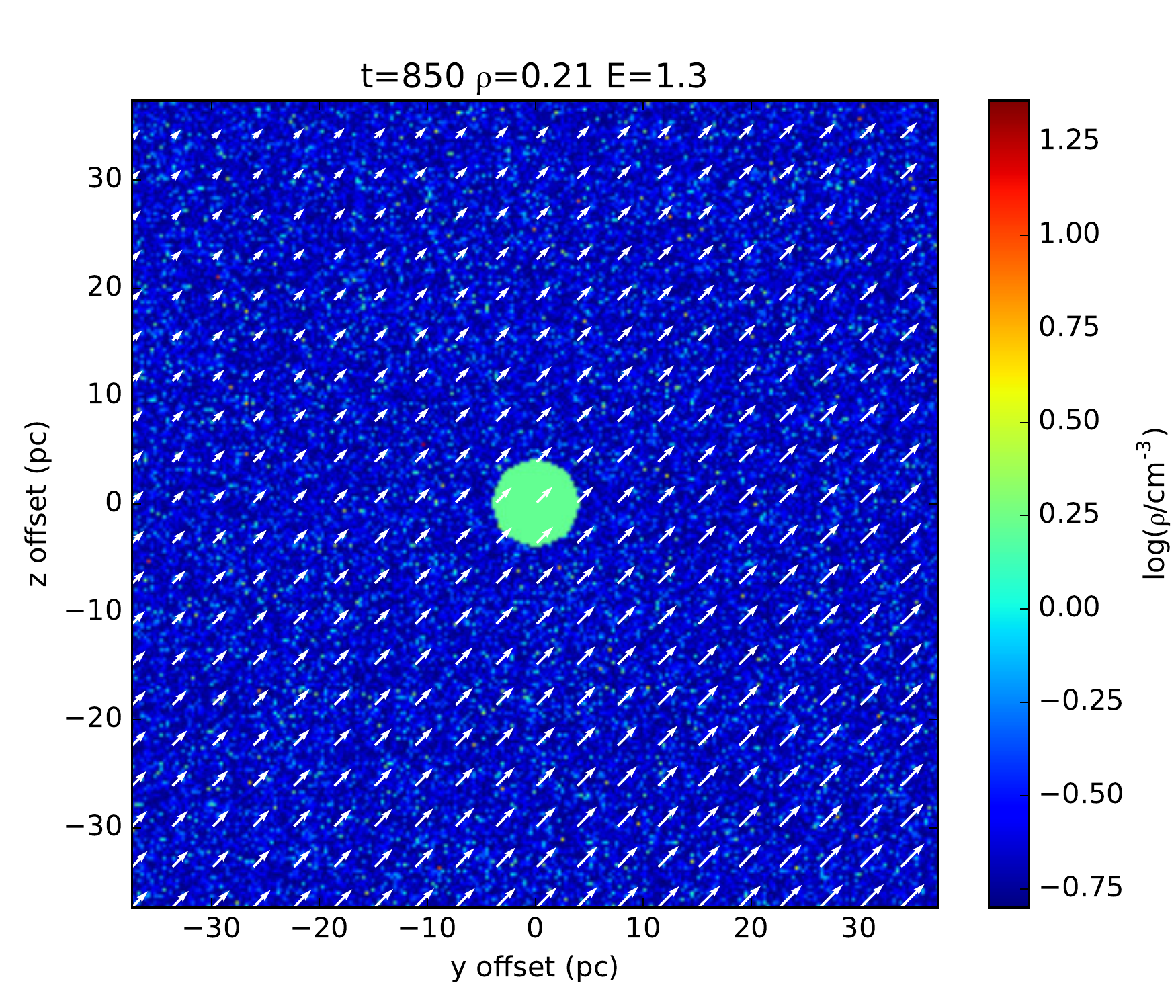}\newline
    \includegraphics[width=0.323\textwidth]{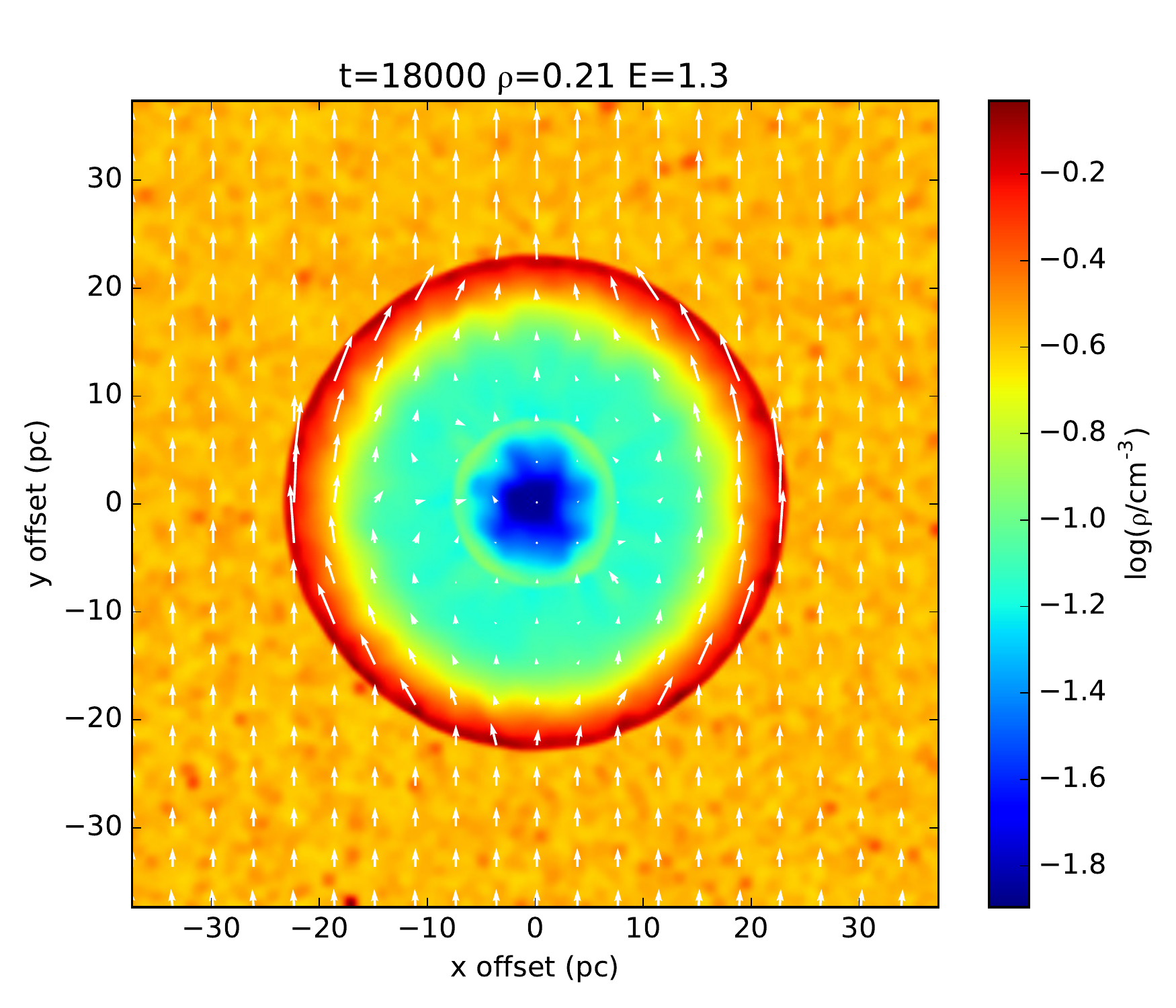}
    \includegraphics[width=0.323\textwidth]{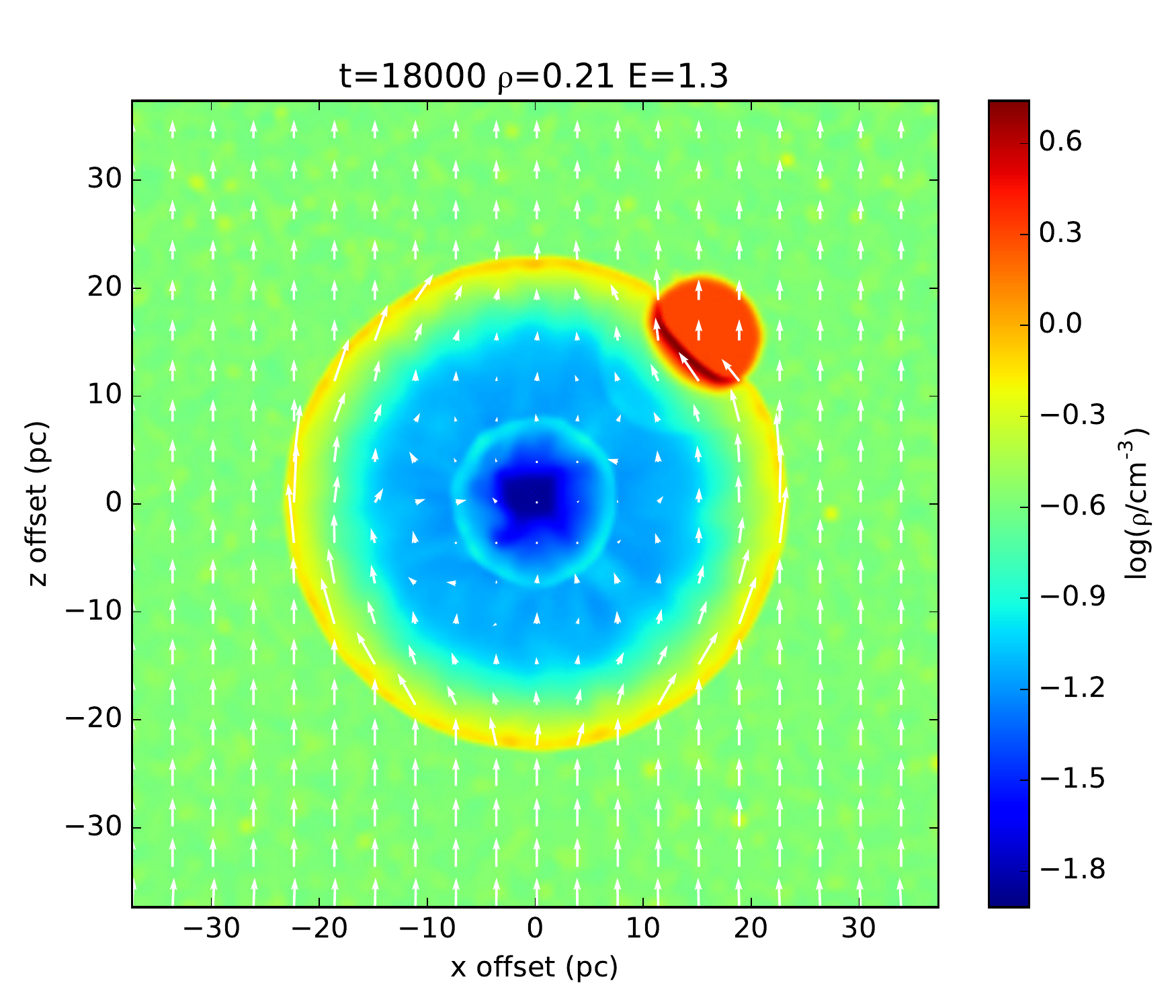}
    \includegraphics[width=0.323\textwidth]{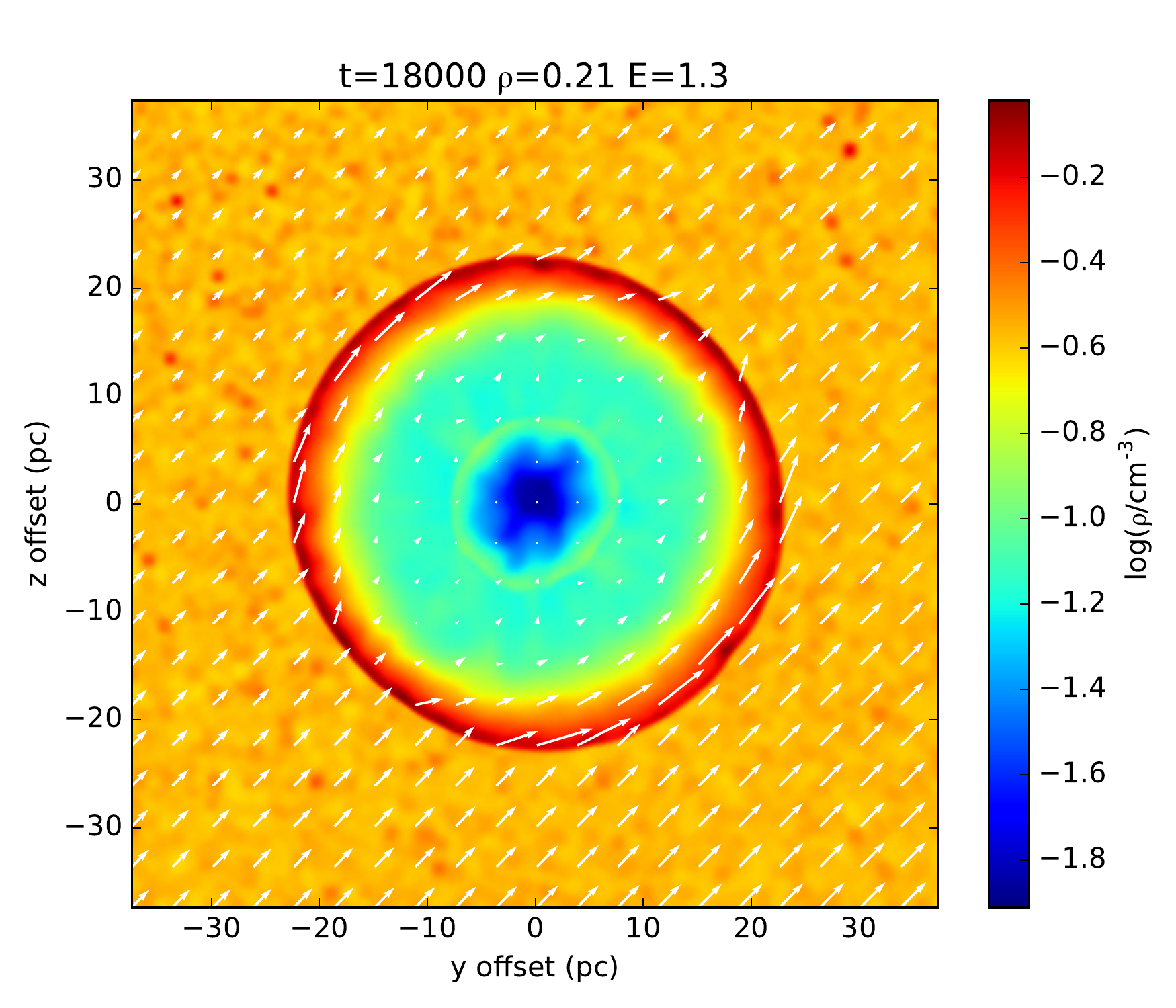}
    \caption{Simulated density-magnetic field images. The images are slices through the center of the box along
    each axis. The background is the density distribution, while the
    white arrows show the direction and intensity of magnetic field. The top panels show the initial conditions
    in the simulation, while the bottom panels show the result after 18000 years. The central magnetic intensity is
    9$\mu G$ in the top panels.}
\label{fig:simulation}
\end{figure*}

We present the simulation result at 18000 years in the bottom panels of Fig.~\ref{fig:simulation}.
In the figure, the direction and length of arrows show the direction of magnetic field and the magnetic intensity,
while the backgrounds show the density.
We can see obvious magnetic amplification both at shock region and interaction region.
In y-z plane, the magnetic intensities on the UL (upper left) and LR (lower right) are larger than their surrounding
regions, and the intensity of the LR is larger than the UL's.
The direction of magnetic field follows the shapes of edges well.
The density and magnetic intensity in the inner SNR become very low, while the outer shell is dense and has stronger
magnetic field.

\begin{figure*}
    \centering
    \includegraphics[width=0.472\textwidth]{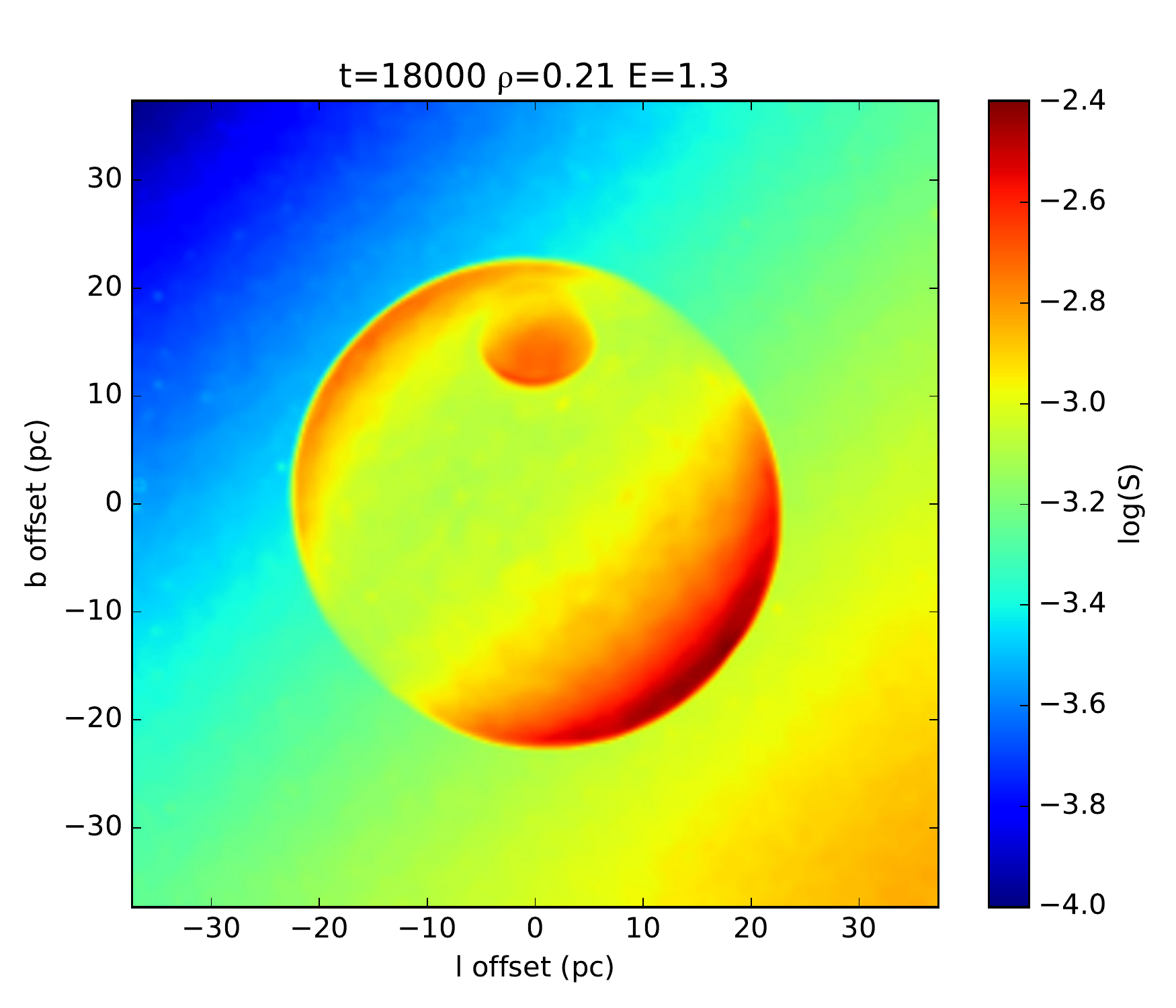}
    \includegraphics[width=0.472\textwidth]{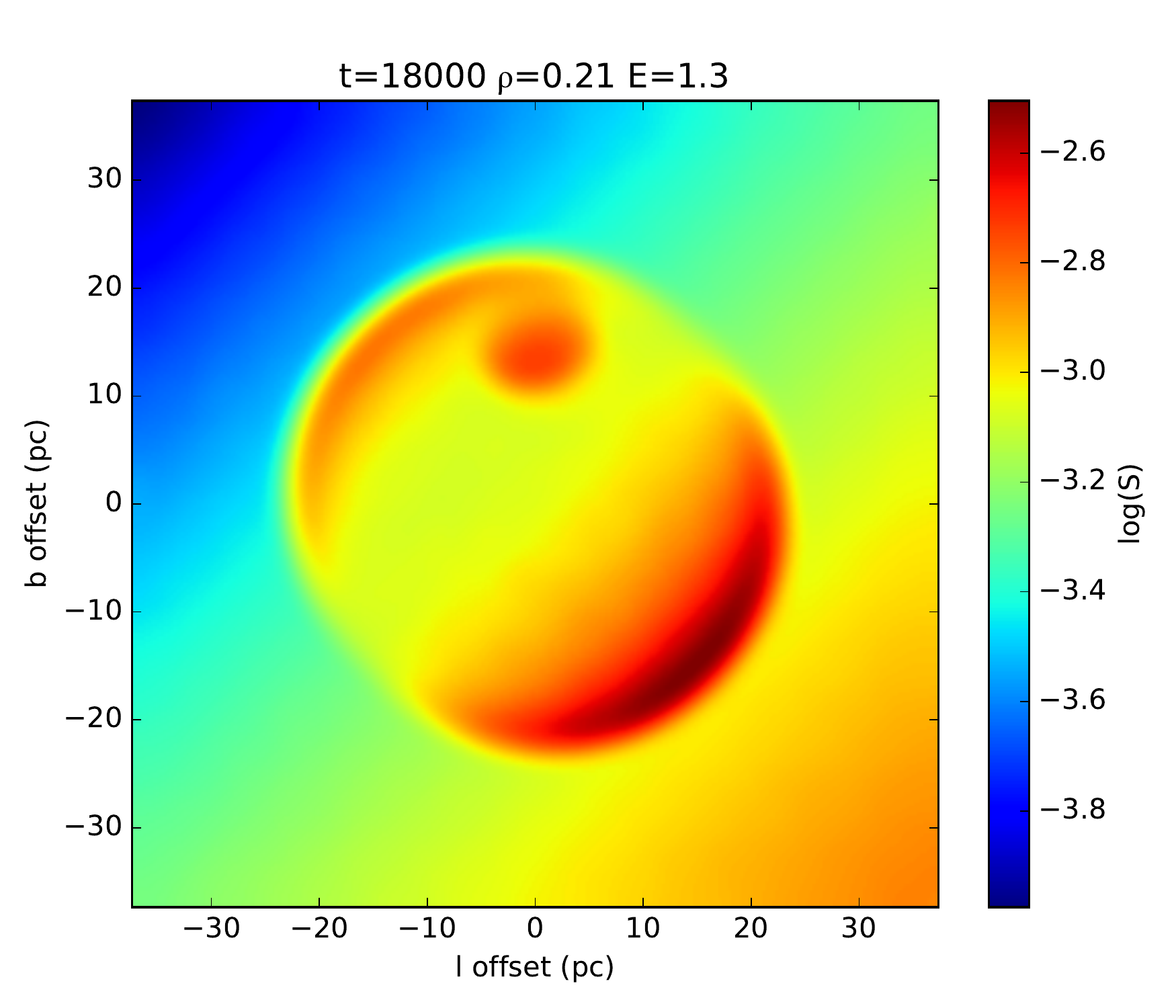}
    \caption{The relative radio flux density after 18000 years for SNR W51C. The right panel results from
    gaussian smoothing with $\sigma$=4. Here 1 pc is equal to 0.8\am assuming a distance of 4.3 kpc.}
\label{fig:flux}
\end{figure*}

Fig.~\ref{fig:flux} shows the final radio flux density maps converted from the density maps.
We test the simulation taking index = 1.0, 3.0. When index = 3.0, the radio image is similar to the image
Fig.~\ref{fig:flux} at index = 2.4.
When index = 1.0, the radio image shows more random components, but other features are similar to the image at index=2.4.
Thus we only present the results for index=2.4.

\begin{figure}
    \centering
    \includegraphics[width=0.472\textwidth]{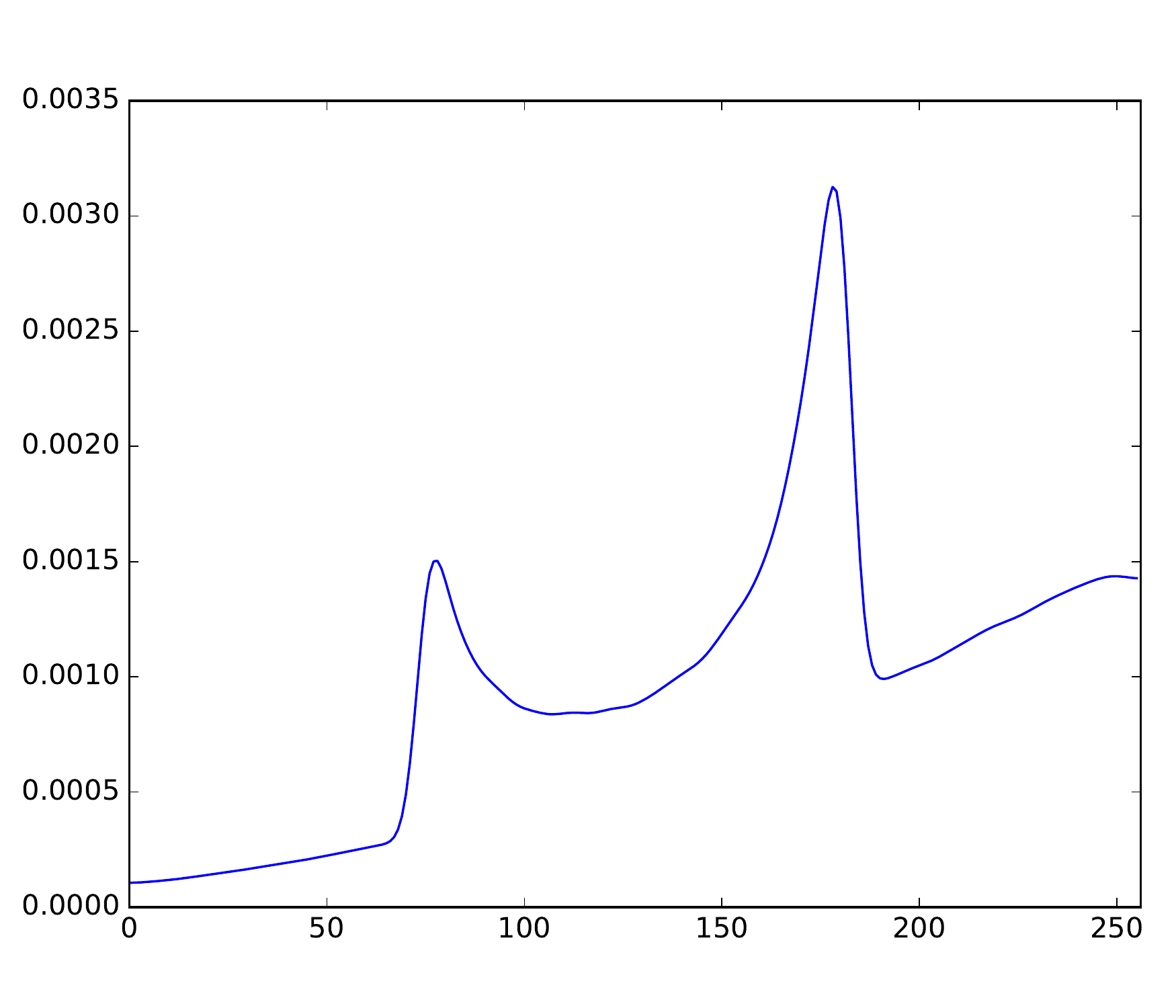}
    \caption{The relative radio flux density along the diagonal line of UL to LR in the right panel of Fig.~\ref{fig:flux}. }
\label{fig:change}
\end{figure}

The right panel uses $\sigma = 4$ for 2D Gaussian smoothing.
The UL edge is weaker but still obvious.
The center is the weakest region in the flux density image, and there is obvious radio emission towards the interaction region.
Without the Gaussian smoothing, the simulation cannot produce the thickness of the real LR edge , even if we take the density
distribution into consideration.
But when $\sigma = 4$, the simulation is consistent with the observation.
Fig.~\ref{fig:change} shows that the UL edge is about 2 times dimmer than the LR edge.

\begin{figure*}
    \centering
    \includegraphics[width=0.472\textwidth]{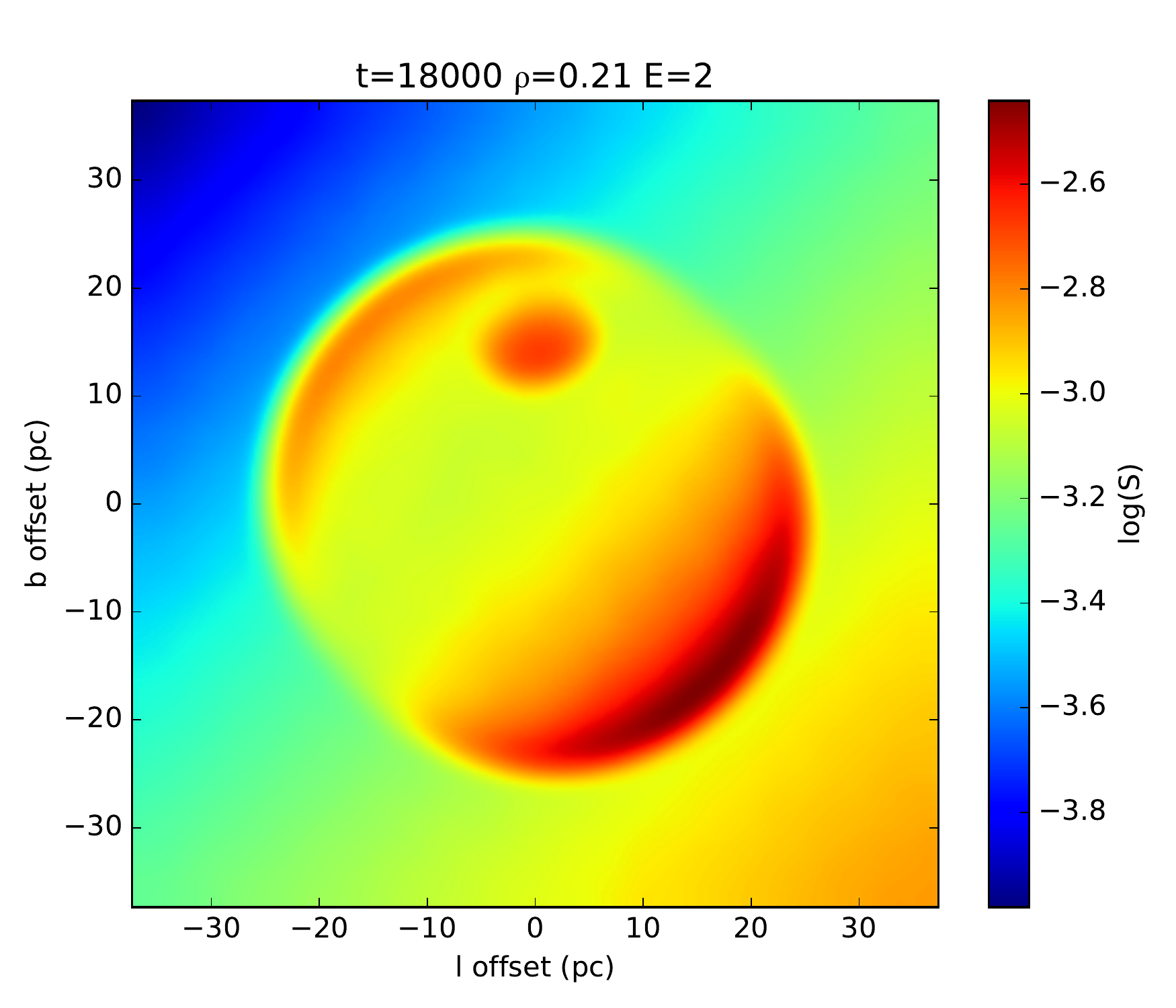}
    \includegraphics[width=0.472\textwidth]{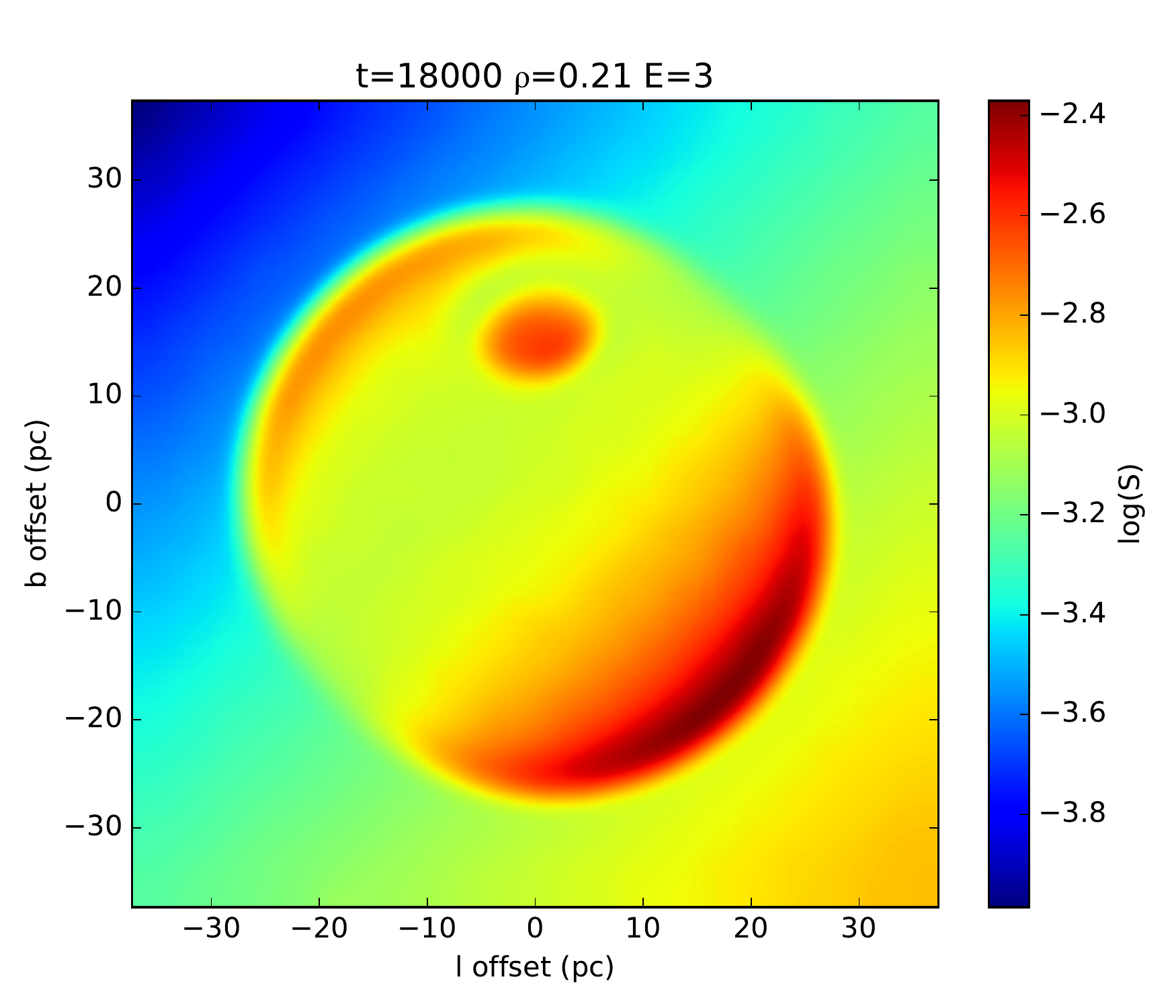}\newline
    \includegraphics[width=0.472\textwidth]{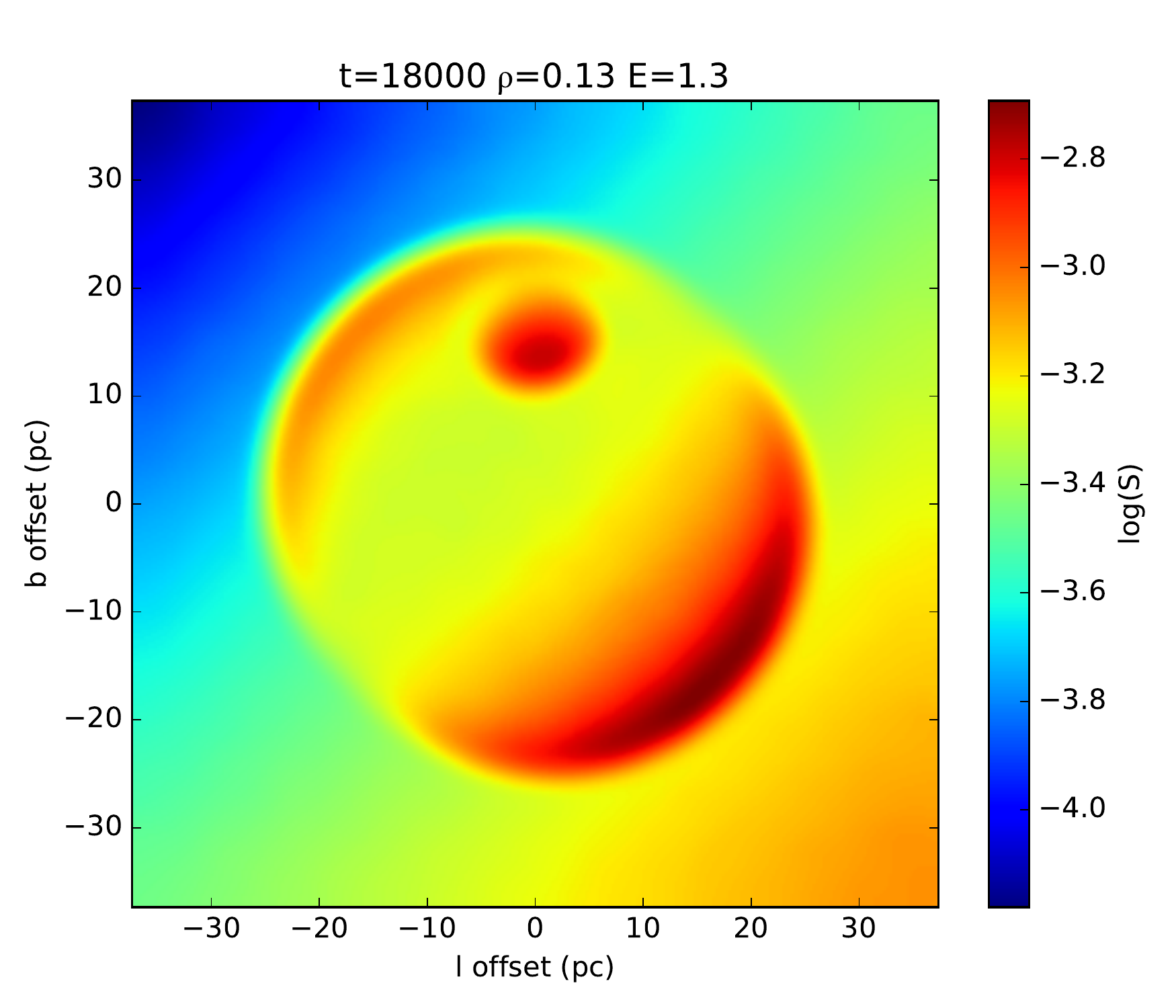}
    \includegraphics[width=0.472\textwidth]{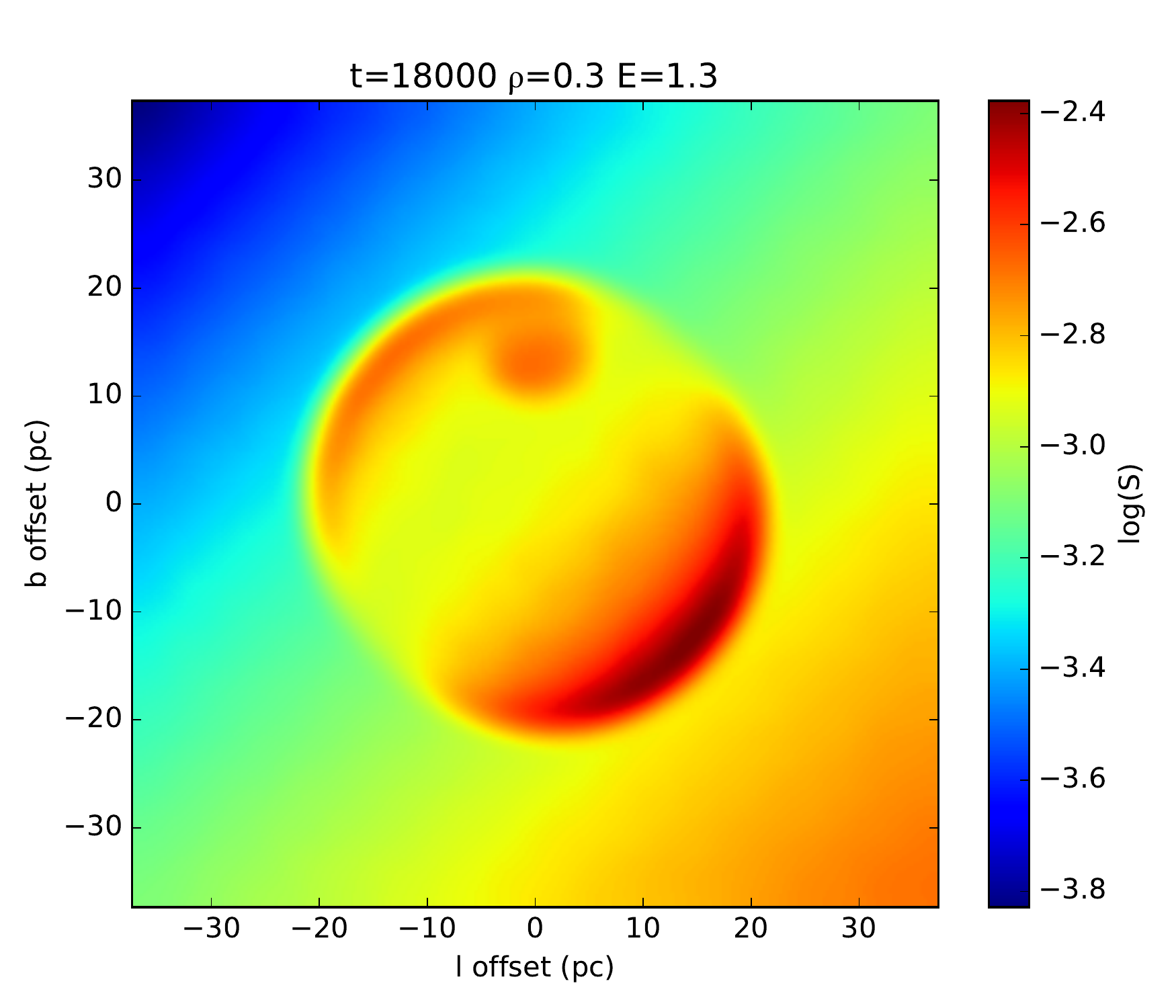}
    \caption{The relative radio flux density in different parameters with $\sigma$=4. The top two panels change the explosion
    kinetic energy to 2.0 and 3.0 $\times$ 10$^{51}$ ergs, while the bottom two panels
    change the density of ISM to 0.13 cm$^{-3}$ and 0.3 cm$^{-3}$. Here 1 pc is equal to 0.8\am assuming a distance of 4.3 kpc.}
\label{fig:moreflux}
\end{figure*}

The results based on different parameters are shown in Fig.~\ref{fig:moreflux}.
The UL edge is obvious in all images.

\subsection{Magnetic field}

\begin{figure}
   \centering
   \includegraphics[width=0.51\textwidth]{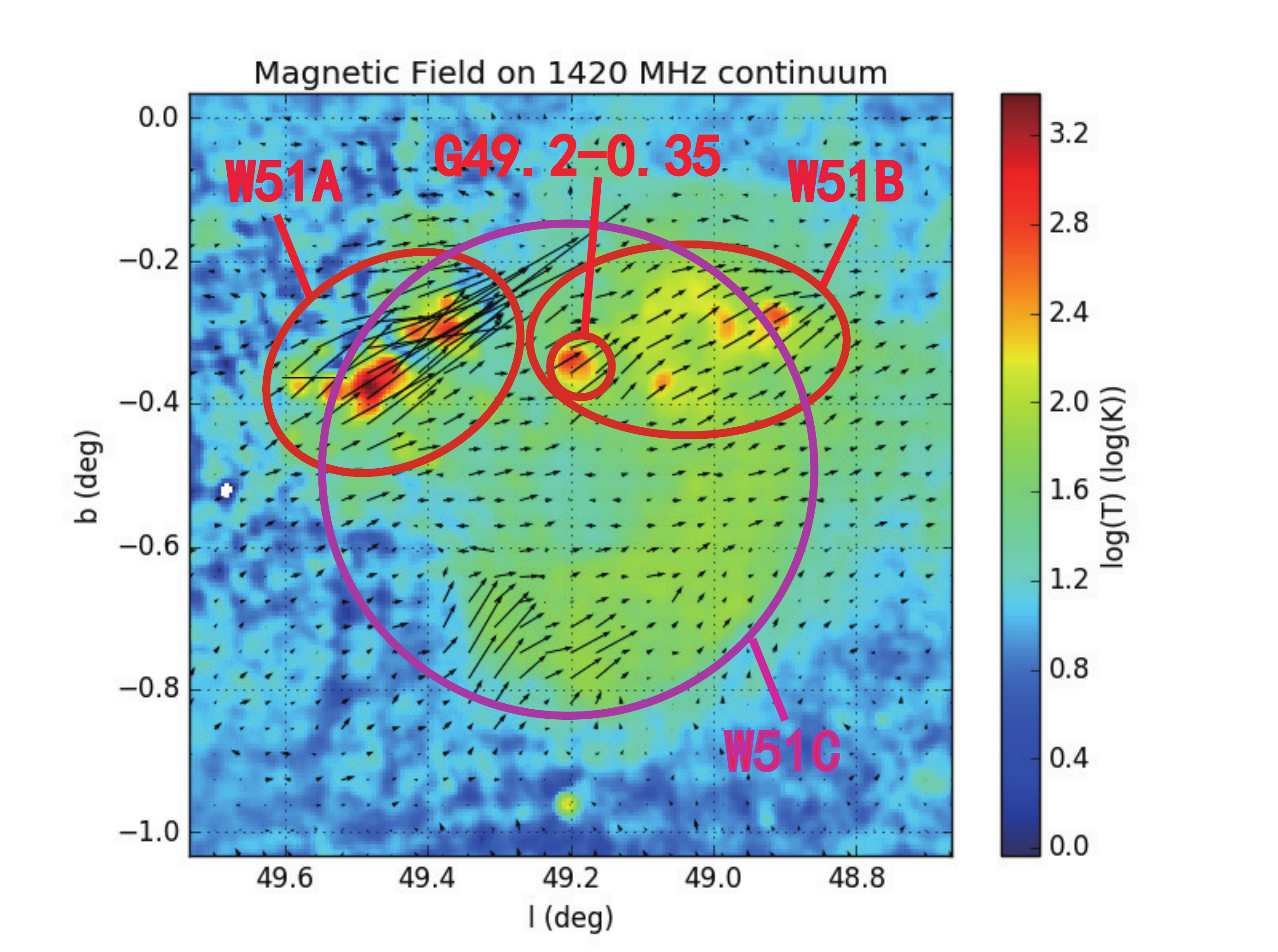}
   \includegraphics[width=0.51\textwidth]{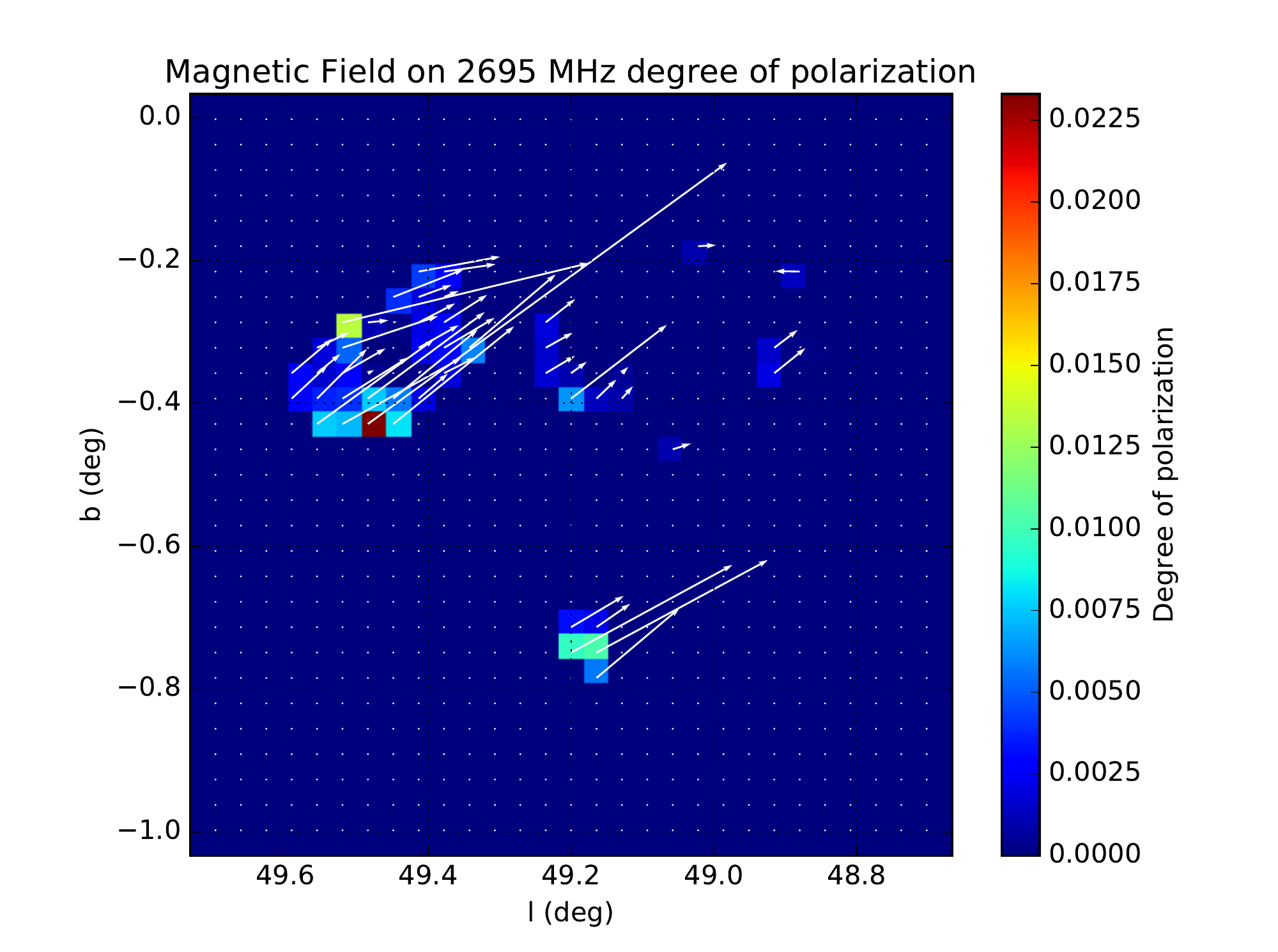}
   \caption{In the top panel, the background is 1.4 GHz continuum image from \textit{VGPS}, while the black arrows show
   the direction of magnetic field. The length of arrows shows the polarization intensity (mK) and the largest intensity is
   1581 mK. In the bottom panel, the background is the degree of polarization on 2695 MHz, while the white arrows show the
   direction of magnetic field. The length of arrows shows the degree of polarization and the largest degree of
   polarization is about $2\%$.}
\label{fig:mag}
\end{figure}

The observed magnetic field structure of W51C is shown in Fig.~\ref{fig:mag}.
There are four regions with strong polarization, the northeast (NE), the middle (M), the northwest (NW), and the south (S).
They all have similar magnetic field direction, even though the direction in NE region is a little chaotic.
The polarization of NE region is the strongest.
After subtracting the instrumental polarization, the polarization in the NW region disappears, while the other three regions
still have polarization.
The direction in NE region becomes more regular, which possibly means we have subtracted enough instrumental polarization.
The NE, M, and S regions respectively overlap with W51A, G49.2-0.35, and W51C well.

\subsection{OH masers and absorption}

\begin{figure}
   \centering
   \includegraphics[width=0.51\textwidth]{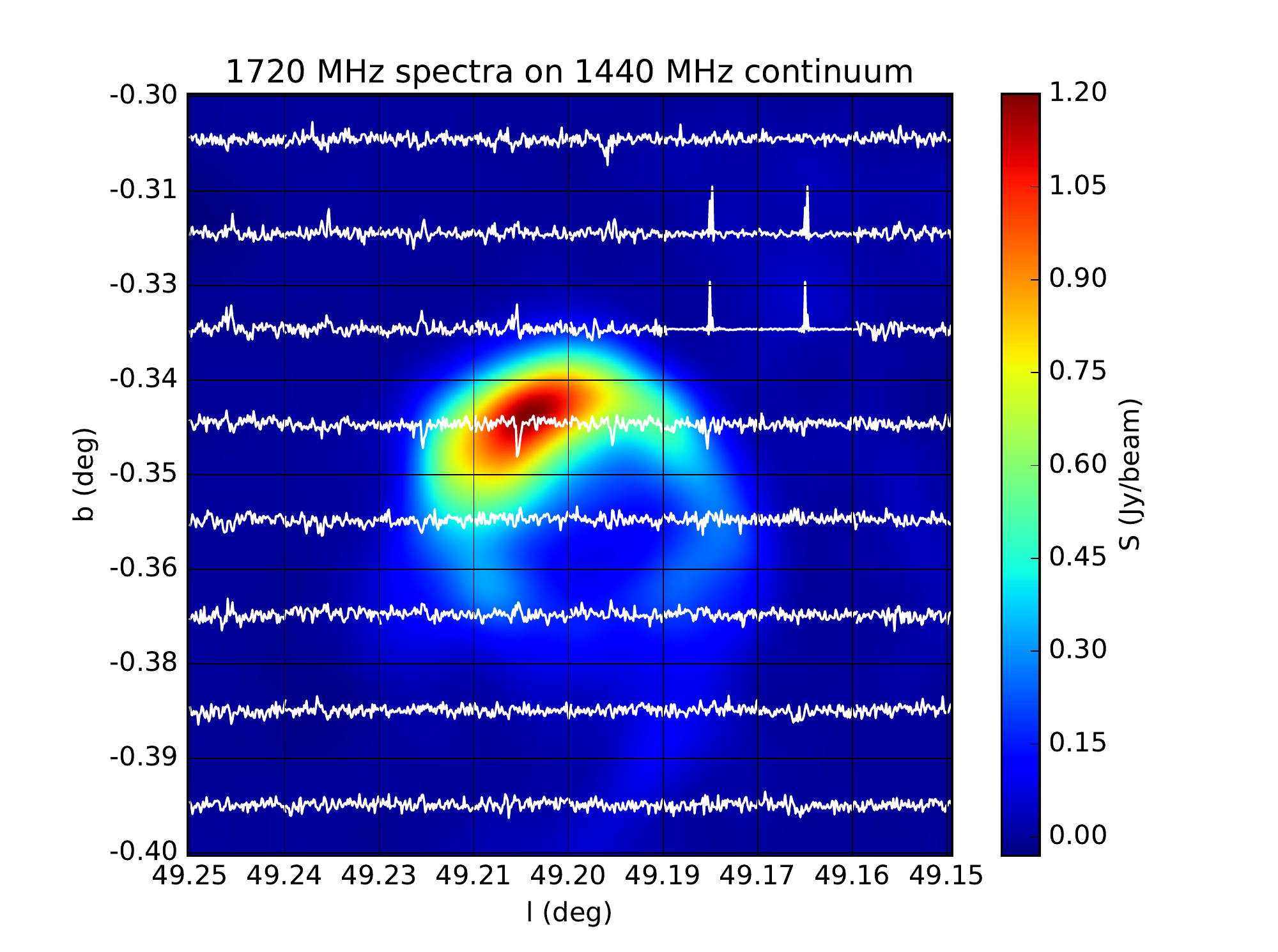}
   \includegraphics[width=0.51\textwidth]{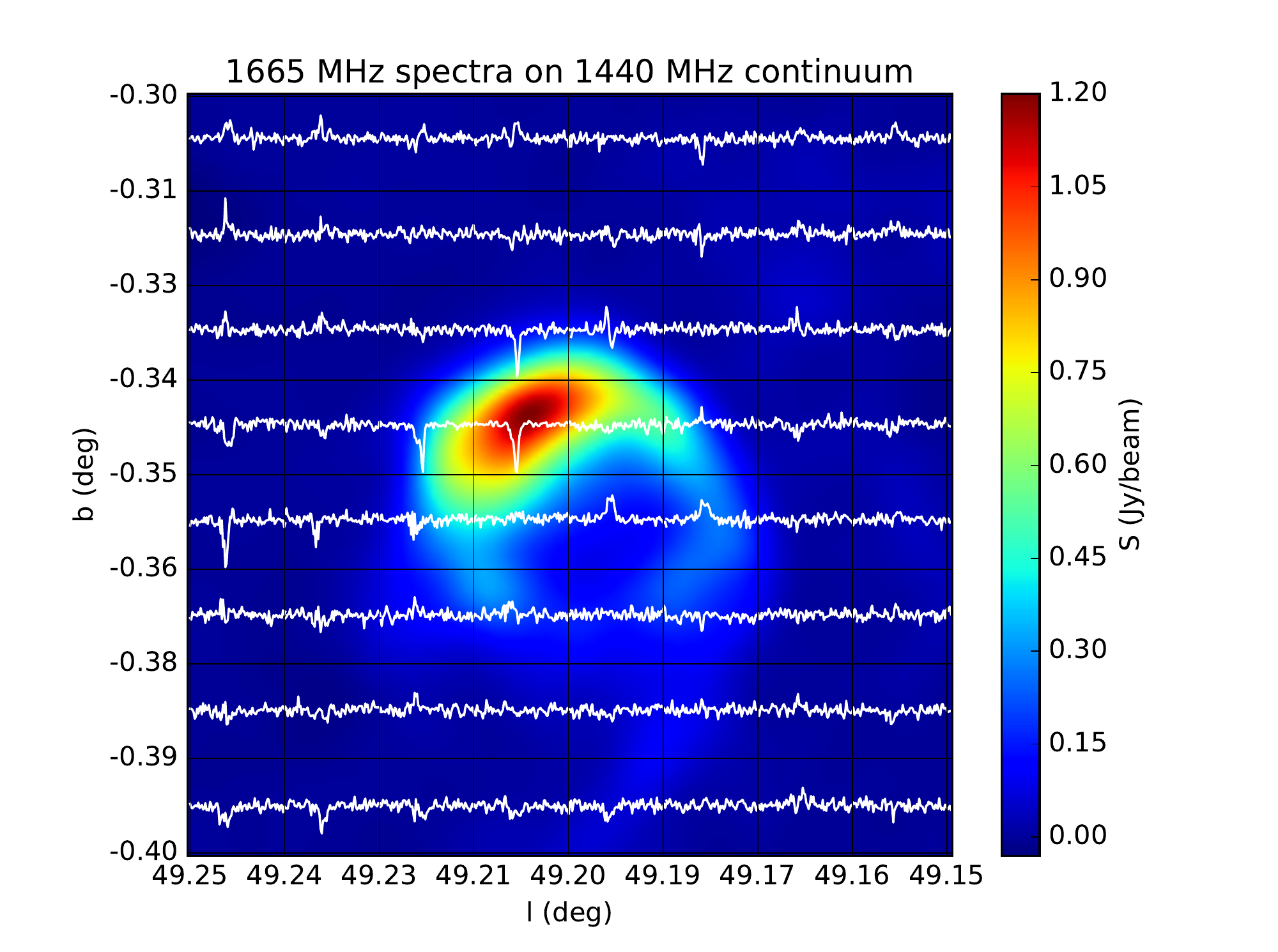}
   \includegraphics[width=0.51\textwidth]{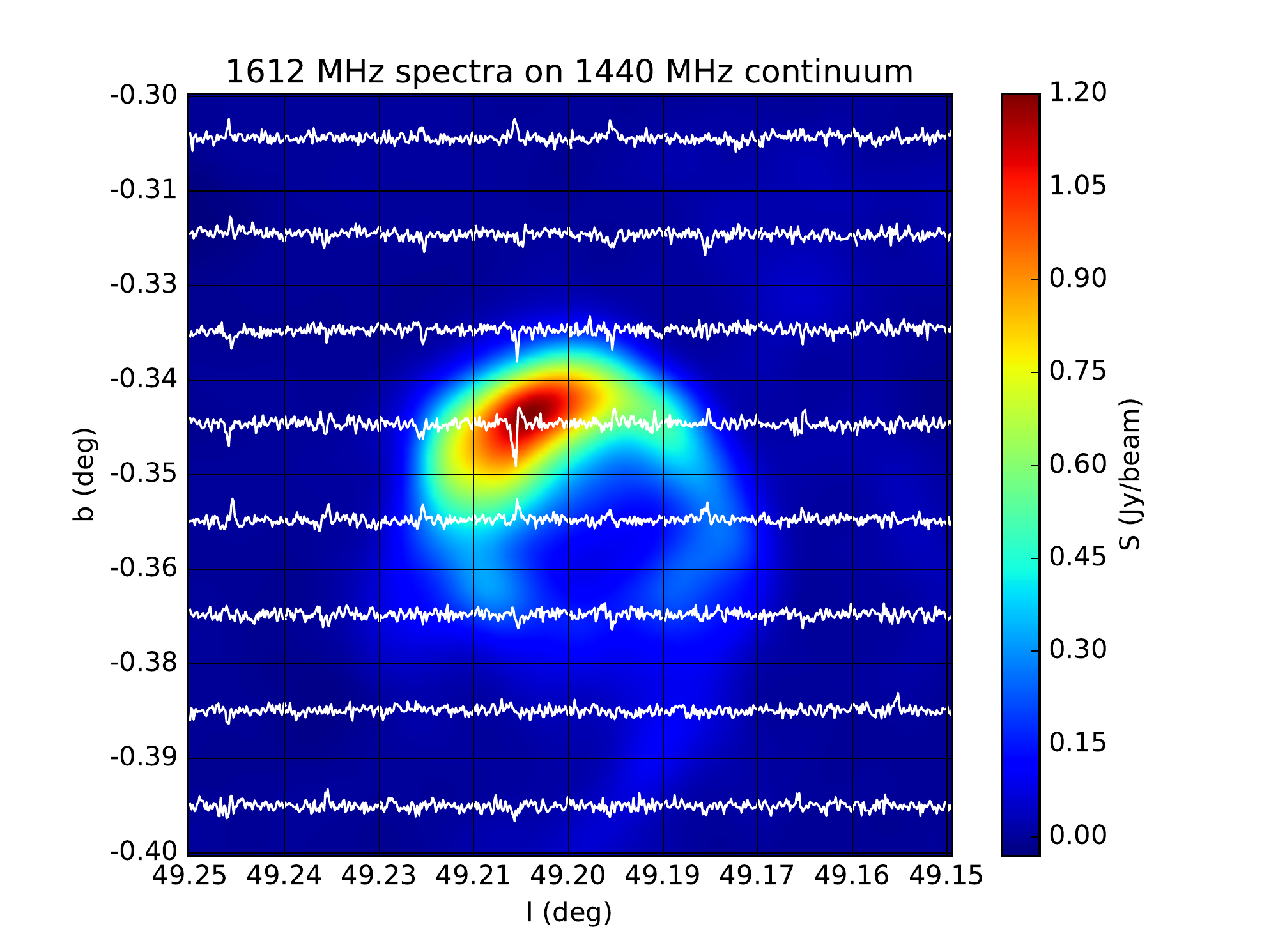}
   \caption{The three figures respectively show OH spectral maps of three frequencies (1720/1665/1612 MHz). The background
   is 1440 MHz continuum image from \textit{THOR}. The spectrum in every black square shows the mean spectrum from -58.5
   \kms to 135 \kms in that region. In the first panel, the scale in the region ($49.16^{\circ}<l<49.19^{\circ}$,
   $-0.34^{\circ}<b<-0.31^{\circ}$) is different from other regions, because the 1720 MHz OH maser in the region is so strong
   that we cannot show it in same scale. In maser regions, the spectral intensities are divided by 20.}
\label{fig:OH}
\end{figure}

There is a high-resolution 1720 MHz observation for the region \citep{2005ASPC..340..334B},
but no high-resolution 1665/1612 MHz observation.
The \textit{THOR} data make it possible to reveal the spatial relation between the 1720 MHz emission and the 1720/1665/1612 MHz absorption.
We present OH spectral maps for the region in Fig.~\ref{fig:OH}.
The background is \textit{THOR} 1.4 GHz continuum image.
The 1720 MHz OH maser is strong and located away from the small \hii region G49.2-0.35 in W51B.
Here we want to indicate that the radio emission of the interaction region in our simulation, is totally different from
\hii region G49.2-0.35.
n fact, we do not detect the radio emission from the interaction region, but the 1720 MHz OH maser and weak polarized radio emission
are good evidence that it exists.
The 1720/1665/1612 MHz OH absorption lines are obvious at the bright region of G49.2-0.35, however,
at the dimmer region, there are some broaden emission lines, in which the 1665 MHz OH emission is stronger.
Meanwhile, some emission and absorption features are far away from the \hii region.
% They are possibly from W51A or W51B.
All the features have similar velocities, so they all possibly originate from the W51 complex.

\begin{figure}
   \centering
   \includegraphics[width=0.51\textwidth]{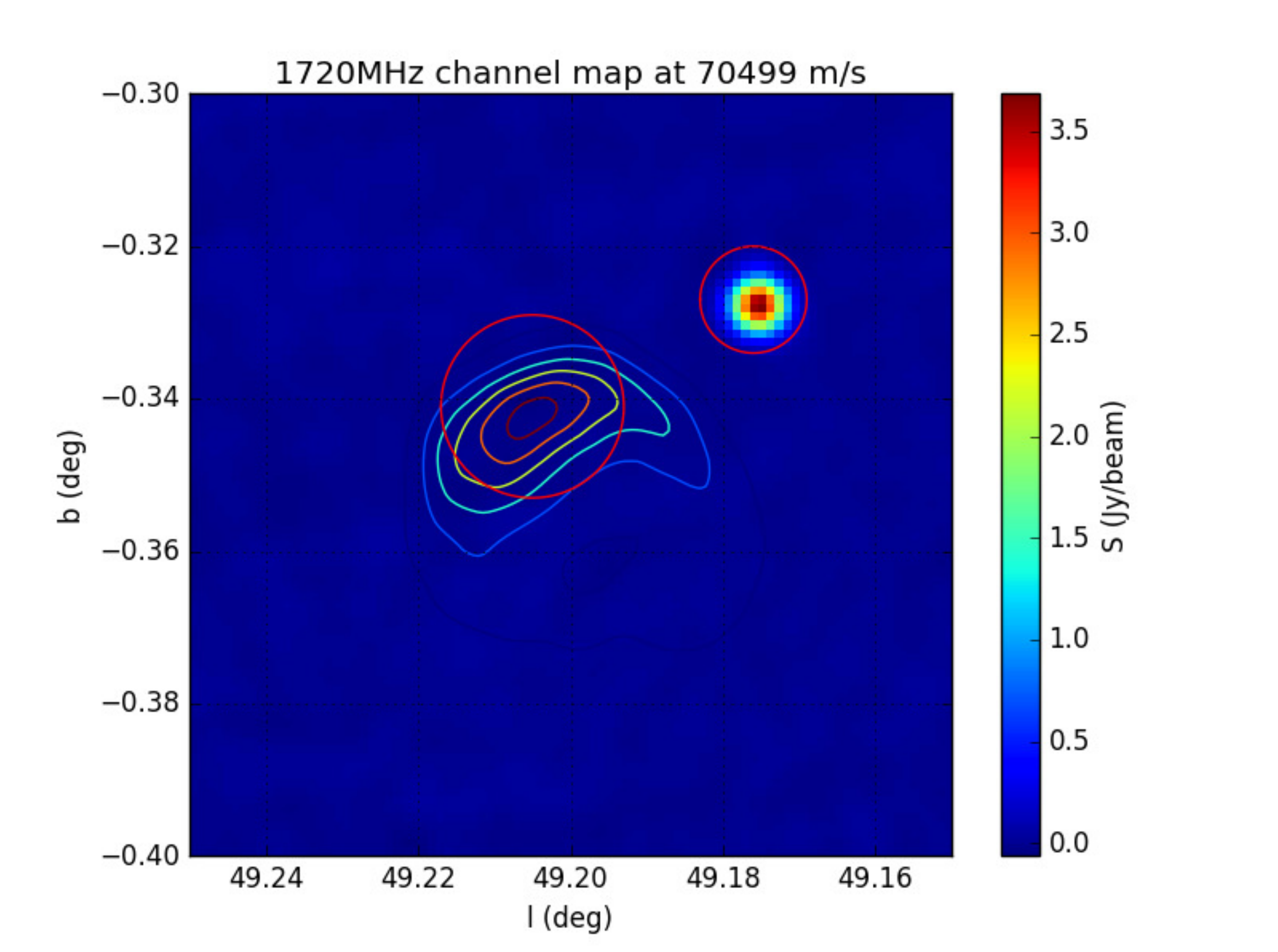}
   \includegraphics[width=0.51\textwidth]{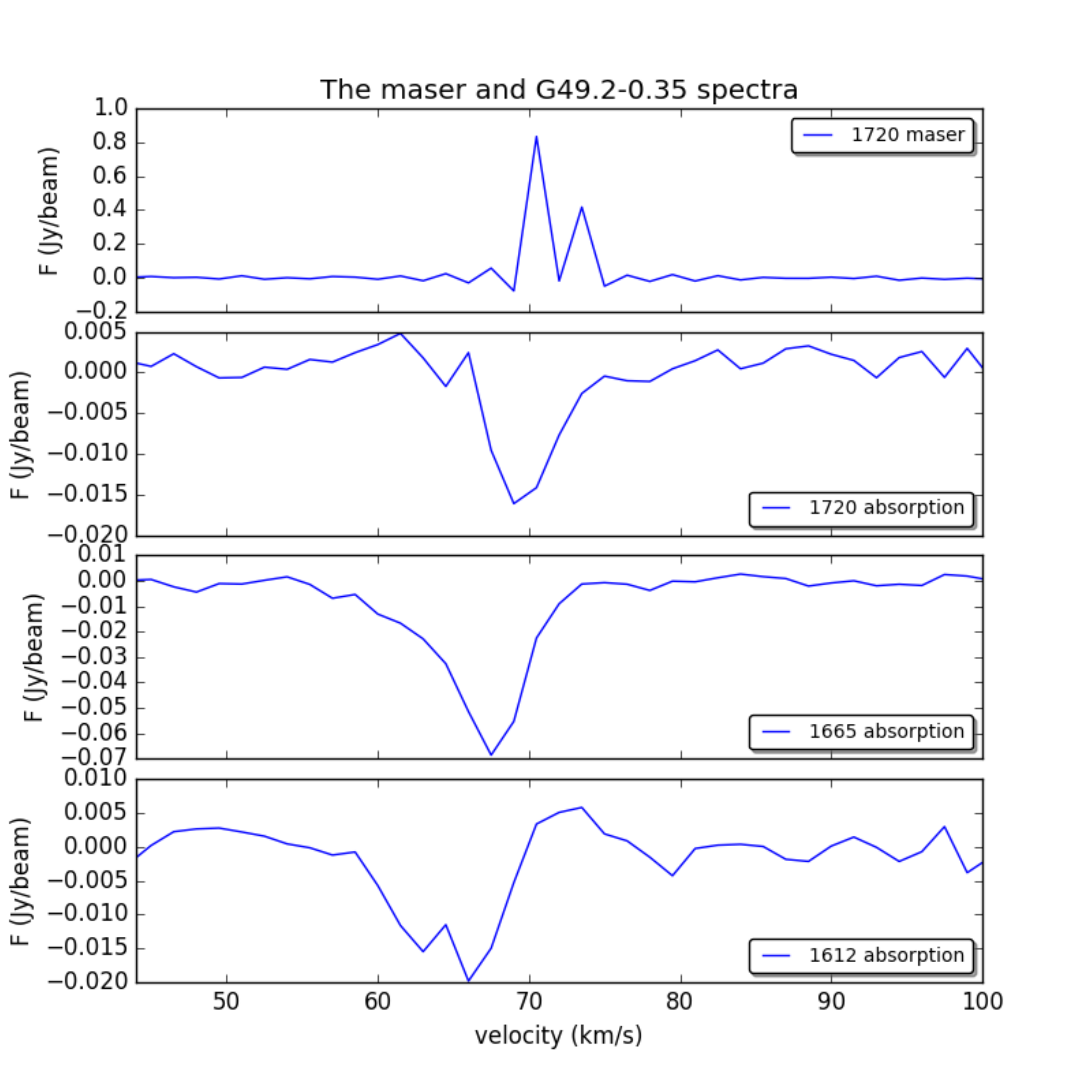}
   \caption{The top panel shows the 1720 MHz channel map towards G49.2-0.35, where the OH maser is very obvious. We plot the
   contours of 1440MHz continuum from \textit{THOR} on the map. The right red circle shows the maser region, while the left
   red circle shows the absorption region. In the bottom panel, we plot the spectra in the two regions. The first spectrum is
   the 1720 MHz OH emission at the maser region, and the rest spectra are the absorption at the \hii region .}
\label{fig:spectra}
\end{figure}

Fig.~\ref{fig:spectra} shows the sites and the spectra of the \hii region G49.2-0.35 and the 1720 MHz OH maser.
In the top panel, the background is \textit{THOR} 1720 MHz channel map at 70.5 \kms, the contours show
\textit{THOR} 1.4 GHz continuum intensities.
The bottom panel presents the spectra of the maser region and \hii region at different frequencies.
The 1720 MHz OH maser is completely separated from the small \hii region G49.2-0.35, but still in the large \hii region W51B.
The 1720 MHz peak and the 1720 MHz absorption feature have a velocity difference of 2 $\sim$ 3 \kms,
the absorption velocities decrease with the decrease of frequencies.

\section{Discussion}
\subsection{The new northeast edge}
\begin{figure*}
   \centering
   \includegraphics[width=0.472\textwidth]{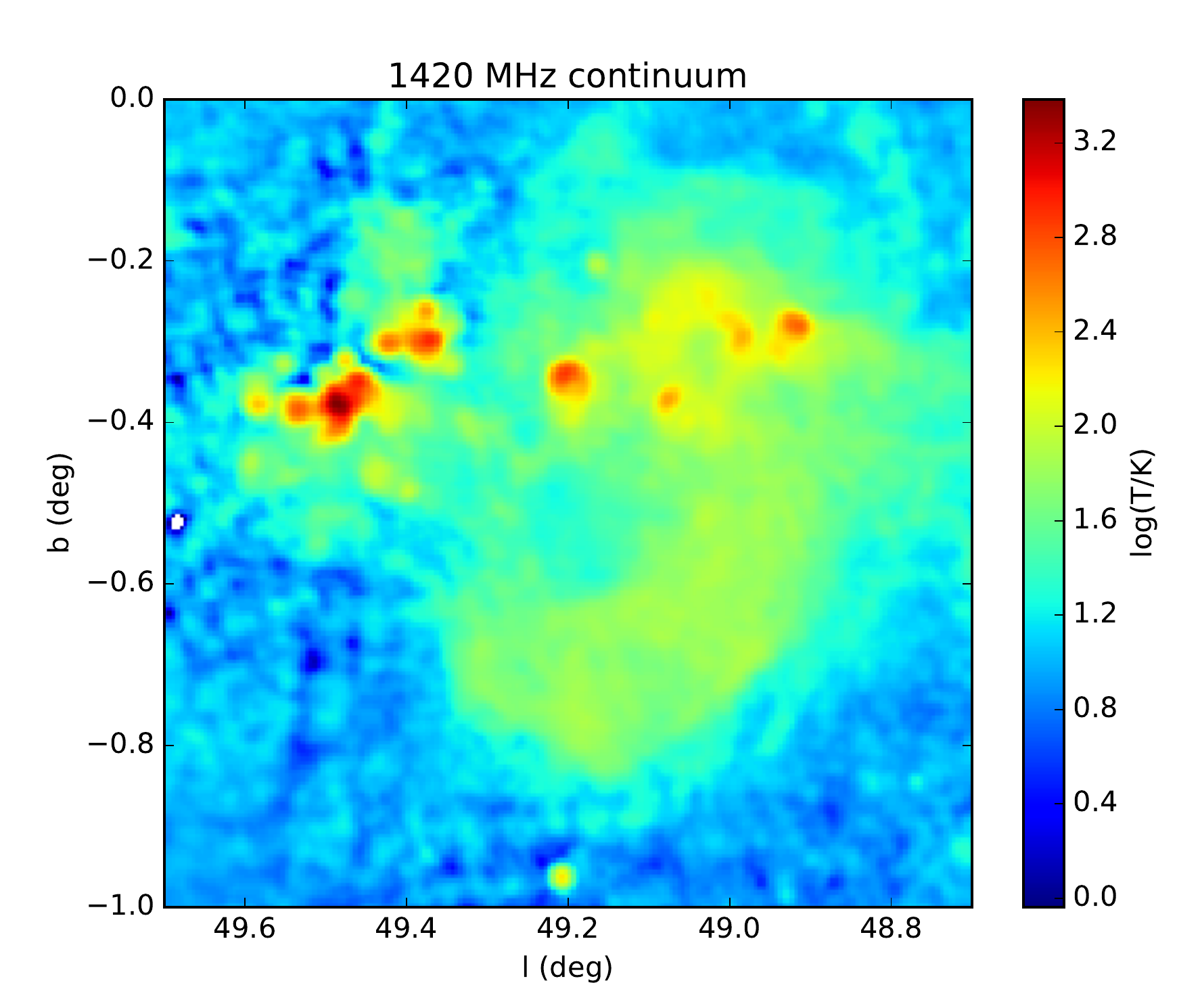}
   \includegraphics[width=0.472\textwidth]{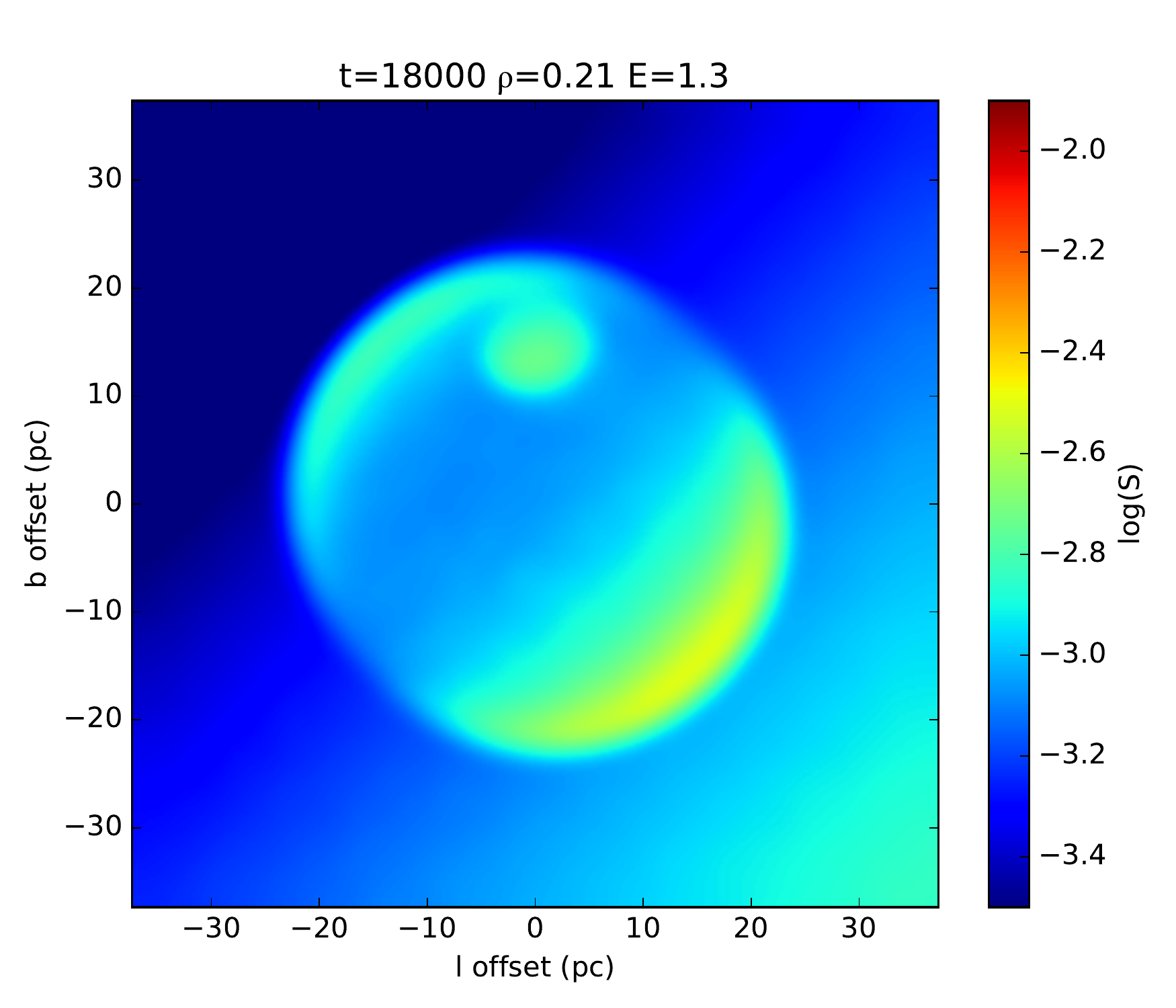}
   \caption{The left panel shows the 1420 MHz continuum map of W51 complex region. The right panel is the result after changing
   the color style of the right panel in Fig.~\ref{fig:flux}. If we take 4.3 kpc as the distance, the two panels have same size.}
\label{fig:compare}
\end{figure*}

To get the semicircular shape, we modify the method used by \citet{Orlando2007}.
We find that the thickness of the shell is related to the density distribution of ISM.
In uniform ISM, the shell is usually thinner \citep{Orlando2007}.
We choose power-law distribution as the initial density distribution.
Our simulation always shows that there appears a north-east edge at the site of W51A.
Because the W51A is so bright that it totally covers the new edge, nobody in previous studies recognized such an edge
(see the comparison between the observation image and the simulation image in Fig.~\ref{fig:compare}).

Based on the simulation result, the new edge is about 2 times dimmer than the southern edge (see Fig.~\ref{fig:change}).

Combining the derivation and some observations, we can predict the flux density of the new edge.
The southwestern edge of W51C has a brightness temperature of about 60 K at 1.4 GHz (see Fig.~\ref{fig:mag}), so the new
northeastern edge should have a brightness temperature of 30 K, larger than the sensitivity of \textit{VGPS}.
Using a spectral index of 0.25, for W51C, we can get the brightness temperature of 14 K at 11 cm, then the new edge should have
a brightness temperature of 7 K at 11 cm.
The degree of polarization of W51C is 3$^{+1.8}_{-1.0}\%$ \citep{1974A&A....32..375V} at 11 cm.
If we believe the degree of polarization of the new edge is the same as the detected edge, its polarization intensity
should be approximately 210 mK, larger than the sensitivity of Effelsberg 11 cm polarization survey.
The magnetic amplification in our simulation is obvious in the NE region, so its polarization flux density should be large.
Generally, the polarization of an \hii region at this wavelength is rare.
If we find polarized light along the LoS of the NE region, it should originate from W51C not W51A.

Fig.~\ref{fig:mag} truly reveals the polarized light towards the NE region.
A part of SNR W51C, i.e. the southwestern region, in which \citet{1974A&A....32..375V} discovered obvious polarization,
does not show any polarization in Fig.~\ref{fig:mag}.
This is possibly because we subtract too much instrumental polarization, which implies the polarization in present image is
possibly the lower limit.
The NE and S regions have same magnetic-field direction, which is similar to our simulation.
Since the polarized emission of the S region is from W51C, the polarization of the NE region is likely also from W51C.
The directions of instrumental polarization \citep{1987A&AS...69..451J} and background polarization \citep{1999A&A...350..447D}
are both different from the direction of observed polarization in this region.
Moreover, there is no other polarized source towards this region \citep{Xu2014}.
Thus, we believe the polarization towards the NE region is mainly from W51C.
In addition, \citet{1974A&A....32..375V} get a total flux density of 51.5 Jy at 11 cm for W51C, so the smallest flux density
of the new edge is about 25.7 Jy.
\citet{1994JKAS...27...81M} mentioned strong non-thermal flux density of 28 Jy at 11 cm towards W51A, but they are not sure
about its origin.
All these clues point to an newly-found northeastern SNR edge.

In fact, every SNR explodes approximately spherically, so most of SNRs should have two edges.
Even an inhomogeneous ISM is hard to yield a one-edge SNR.

% Considering the relative velocities between progenitor stars and the ISM, we can get a better one-shell shape, but the
% shape will be a little strange, just like the Cygnus Loop \citep{Fang2017}.
% If the background emission of the host galaxy is strong or the sensitivities of telescopes are very low, a two-shells SNR
% will also be identified as a one-shell SNR.
% This will lead to some severe misunderstandings.
% Thus, we suggest to further identify all one-shell SNRs detected so far.
\subsection{The interaction region}
\citet{1994JKAS...27...81M} estimated a probable non-thermal flux density of 9 Jy at 11 cm towards \hii region G49.2-0.35,
which is further supported by \citet{Brogan2013}.
We also find weak polarized emission towards it in Fig.~\ref{fig:mag}.
Similar to the new edge, this possibly originates from the SNR.
This region is next to the explosion center along the LoS, so this non-thermal emission is possibly from the shell expanding
along the LoS.
Due to the projection effect, usually a shell expanding along the LoS has low flux density and is hard to be detected.
However, such a shell will be detectable if the SNR interacts with a MC (see Fig.~\ref{fig:flux}).
In fact, we detect the 1720 MHz OH maser in the region, which favors such an SNR-MCs interaction in the region.

In addition, shocks propagating along the LoS will not change the initial direction of magnetic field towards us.
The observed direction of magnetic field is same as our assumed initial direction of magnetic field in the region.
And the direction of the Galactic large-scale magnetic field around W51 complex is different from the direction in this region.
So our assumptions on the origin of the non-thermal emission at this region and the initial direction of
magnetic field are self-consistent.

The synthesis of OH molecules is a totally different proceeding from the generation of masers.
Shocks dissociate the molecules in MCs and produce OH molecules behind the shock \citep{Wardle1999}.
Masers are generated with strict conditions: the temperature, density, and OH column density must be all suitable.
Thus most of OH molecules cannot form masers.
These molecules may absorb the continuum emission from strong background source, such as SNRs and \hii regions, then
generate absorption lines.
\citet{Hewitt2008} believed that detecting a narrow 1720 MHz OH emission line and broadened 1667/1665/1612 MHz absorption lines
in the same region are strong evidence of interaction between SNRs and MCs.
Such absorption will be rare if the background source is weak.
% However, detecting 1720 MHz OH maser is enough to prove the interaction between shocks and MCs \citep{Elitzur1976}.
Fig.~\ref{fig:OH} shows that the 1720/1665/1612 MHz absorption are mainly detected in the region with strong
background continuum emission.
Around the absorption region, there appear emissions in the region where the background continuum emission is weak.
These features are away from the 1720 MHz OH maser.
Since \hii regions can also produce OH molecules, these features are likely related to the \hii region G49.2-0.35.
% This disagrees with the conclusion of \citet{Hewitt2008}.

Fig.~\ref{fig:spectra} reveals that the velocity of the OH absorption features decrease with the frequency of the
OH transition.
The velocity shift may be very local, rather than being an indicator of galactic kinematic distance, but is not related
to the SNR W51C.
The fact of detecting the absorptions against the \hii region G49.2-0.35 implies the \hii region should lie behind the OH clouds producing the
absorption features.
Referring to the studies of \citet{Brogan2013} and \citet{Ginsburg2015}, G49.2-0.35 lies in front of W51C.
If so, the SNR has nothing to do with the OH clouds where these OH absorptions are produced.
% The 1720 MHz OH maser shows very narrow but strong spectral lines.
% The absorption features have velocities lower than the masers, which implies the SNR should lie behind the
% OH clouds producing the absorption features.
% \textbf{However, there are two small problems: (1)the tangent velocity along this LoS is 62 \kms, but observed OH absorption and emission
% velocities are greater than the tangent velocity; (2)OH masers come from small region and likely have significant peculiar motion.
% We think the absolute distance-velocity relation based on the Galactic orbital motion is no longer meaningful due to the large velocities,
% but it is still reasonable to believe larger velocity means larger distance.
% Moreover, although OH masers possibly do not share systemic velocities of their host remnants \citep{Claussen1997}, this situation is not
% serious for W51C, because the velocities of OH masers are similar to the systemic velocity derived from HI absorption/emission
% features \citep{Tian2013}.
% It is difficult to get a clear physical relation between G49.2-0.35 and W51C only based on the emission and absorption features of masers.
% Together with the studies of \citet{Brogan2013} and \citet{Ginsburg2015}, our conclusion is only another possible evidence supporting that
% G49.2-0.35 lies in front of W51C.}
In fact, there are three masers in the region \citep{Brogan2013}, but the spatial resolution of \textit{THOR} is not
enough to distinguish them.
Two of the three are close each other and have almost same velocity.
Thus we can see two lines with peak velocities of 71 and 73 \kms respectively in Fig.~\ref{fig:spectra}.
In addition, 1720 MHz OH masers should be distributed over the SNR, if one SNR shocks the MCs \citep{Wardle2012}.
The fact that there are only three masers in the W51B/C region hints that the contact area of SNR shocks and MCs is likely small.
A small contact area means a small interaction region, so there are only few masers.

In conclusion, we suggest a shock propagating along the LoS, is interacting with a MC.
Such a situation is supported by the polarized emission, the non-thermal emission, and OH spectra.
It can explain the assumed initial magnetic direction and the few OH masers.

\section{Summary}

% \begin{itemize}
We simulate the evolution of SNR W51C from 850 to 18000 years.
Our simulation reveals a new edge, and that its flux density is higher than the sensitivity of \textit{VGPS}.
By analyzing observation data, we find the polarized emission from the new edge, so we conclude that a new edge
does exist.
% \citet{1994JKAS...27...81M} estimated strong non-thermal emission in same region.
% The three evidences prove the new shell likely exists.
In addition, we detect obvious polarized emission in the small \hii region G49.2-0.35 next to the explosion center of simulation, which
can be explained as the result of interaction between SNR shocks and ISM along the LoS.
We find the 1612/1665/1720 MHz OH absorption features against \hii region G49.2-0.3 and the 1720 MHz OH maser emission spectrum at
the W51B/C interface region.
We conclude that the absorptions are mostly from the OH gas in front of the \hii region G49.2-0.35 and not related to the SNR.
% The features of these spectra imply that the \hii region G49.2-0.35 lies in front of SNR W51C.

% \end{itemize}

\acknowledgements
We thank Dr.Reich for his explaining the polarization data of Effelsberg.
Drs. X.Y.Gao and J.Xu also provide suggestions about polarization data.
We thank Drs. Rugel and Beuther for their helpful discussions on \textit{THOR} data.
We also thank Dr.J.Fang for his suggestion about supernova remnants simulation.
We acknowledge support from the NSFC (11473038,11603039).

\bibliographystyle{aasjournal}
\bibliography{mydb}
\end{document}